\documentclass[fleqn,usenatbib]{mnras}

\usepackage{newtxtext,newtxmath}
\usepackage[T1]{fontenc}

\DeclareRobustCommand{\VAN}[3]{#2}
\let\VANthebibliography\thebibliography
\def\thebibliography{\DeclareRobustCommand{\VAN}[3]{##3}\VANthebibliography}

\newcommand{\hii}{\mbox{H{\sc ii}~}}
\usepackage{graphicx}	% Including figure files
\usepackage[skip=0.333\baselineskip]{caption}
\usepackage{subcaption}
\newcommand{\GG}[1]{}
\title[Disk evolution in the twin clusters of W5]{Twins in Diversity: Understanding circumstellar disk evolution in the twin clusters of W5 complex}

\author[Belinda Damian]{
Belinda Damian$^{1,2}$\thanks{belin$\_$93@yahoo.in},
Jessy Jose$^{3}$\thanks{jessyvjose1@gmail.com},
Swagat R. Das$^{3,4}$, 
Saumya Gupta$^{3}$, 
Vignesh Vaikundaraman$^{3,5}$, 
\newauthor
D. K. Ojha$^{6}$,
Sreeja S. Kartha$^{2}$, 
Neelam Panwar$^{7}$ and 
Chakali Eswaraiah$^{3}$
\\
$^{1}$SUPA, School of Physics \& Astronomy, University of St Andrews, North Haugh, St Andrews, KY16 9SS, UK\\
$^{2}$Department of Physics and Electronics, CHRIST (Deemed to be University), Hosur Road, Bengaluru 560029, India\\
$^{3}$Department of Physics, Indian Institute of Science Education and Research (IISER) Tirupati, Yerpedu, Tirupati, Andhra Pradesh, 517619, India\\
$^{4}$Departamento de Astronom{\'i}a, Universidad de Chile, Las Condes, 7591245 Santiago, Chile\\
$^{5}$Max Planck Institute for Solar System Research, Justus-von-Liebig Weg 3, 37077, G\"{o}ttingen\\
$^{6}$Department of Astronomy and Astrophysics, Tata Institute of Fundamental Research, Mumbai 400005, India\\
$^{7}$Aryabhatta Research Institute of Observational Sciences (ARIES), Manora Peak, Nainital 263001, India}

\date{Accepted XXX. Received YYY; in original form ZZZ}

\pubyear{2015}

\begin{document}
\label{firstpage}
\pagerange{\pageref{firstpage}--\pageref{lastpage}}
\maketitle

\begin{abstract}
Young star-forming regions in massive environments are ideal test beds to study the influence of surroundings on the evolution of disks around low-mass stars. We explore two distant young clusters, IC 1848-East and West located in the massive W5 complex. These clusters are unique due to their similar (distance, age, and extinction) yet distinct (stellar density and FUV radiation fields) physical properties. We use deep multi-band photometry in optical, near-IR, and mid-IR wavelengths complete down to the substellar limit in at least five bands. We trace the spectral energy distribution of the sources to identify the young pre-main sequence members in the region and derive their physical parameters. The disk fraction for the East and West clusters down to 0.1 M$_\odot$ was found to be $\sim$27$\pm$2\% (N$_\mathrm{disk}$=184, N$_\mathrm{diskless}$=492) and $\sim$17$\pm$1\% (N$_\mathrm{disk}$=173, N$_\mathrm{diskless}$=814), respectively. While no spatial variation in the disk fraction is observed, these values are lower than those in other nearby young clusters. Investigating the cause of this decrease, we find a correlation with the intense feedback from massive stars throughout the cluster area. We also identified the disk sources undergoing accretion and observed the mass accretion rates to exhibit a positive linear relationship with the stellar host mass and an inverse relationship with stellar age. Our findings suggest that the environment significantly influences the dissipation of disks in both clusters. These distant clusters, characterized by their unique attributes, can serve as templates for future studies in outer galaxy regions, offering insights into the influence of feedback mechanisms on star and planetary formation.

\end{abstract}


%%%%%%%%%%%%%%%%%%%%%%%%%%%%%%%%%%%%%%%%%%%%%%%%%%

%%%%%%%%%%%%%%%%% BODY OF PAPER %%%%%%%%%%%%%%%%%%

\section{Introduction}
\label{sec:intro}
In the contemporary framework of star formation, the growth of low-mass stars is an intricate process that unfolds over time through the accumulation of matter from the circumstellar disks \citep{lynden1974}. Circumstellar disks are a ubiquitous by-product of the star formation process and are hosted by a majority of the low-mass stars at an age of $\sim$1 Myr. They are the birth sites of planets and provide raw materials for planet formation. Hence, their evolution and dissipation timescales are symbiotic with the evolution of planetary systems \citep{morbidelli2016}. The initial stages of star and disk growth are shaped by the dynamic interplay of various mechanisms, notably, the accumulation of disk material onto the star, as extensively discussed in reviews by \citet{bouvier2007} and \citet{hartmann2016}, the expulsion of matter from the disks, achieved through the generation of winds and outflows (\citealt{frank2014,ercolano2017,winter2022}), and influences by localized environmental factors, as elucidated in studies such as those by \citet{pfalzner2005}, \citet{anderson2013}, \citet{fischer2017} and \citet{winter2018}. Each of these processes influences the disk dispersal and in turn the formation of planets (see review by \citealt{manara2023}). 

In this study, we focus on the impact of massive stars on their immediate environment, while also delving into the measurements of disk accretion rates and their correlations with the age and mass of the hosting star. Environment plays a crucial role in the formation and dissipation of disks. Observationally, the influence of the environment has been quantified by the fraction of stars that host a disk in different cluster populations across ages (\citealt{haisch2001,richert2018}). The disk frequency is high at $\sim$ 40-80\% in very young clusters of a few million years and the occurrence of disk decreases gradually in older populations ($>$10 Myr) to $<$5\% (\citealt{sicilia2006,mamajek2009}). The role of environment in disk evolution can be attributed to close encounters of stars in dense regions due to mutual gravitational interaction which can disperse the disk material. Although this interaction disperses the disk, it has been reported to be a secondary effect compared to external photoevaporation (\citealt{scally2001,winter2018}). \citet{winter2018} statistically reported that the local cluster density is required to be above 10$^4$ pc$^{-3}$ to cause significant truncation of the disk in 3 Myr. This impact is negligible in completely disrupting the disk in most of the Milky Way clusters \citep{steihausen2014} but causes significant destructive encounters in very high density regions like Arches cluster \citep{olczak2012}. Whereas, strong UV radiation from massive stars can dissociate and ionize hydrogen molecules and atoms driving external photoevaporation. Due to the erosion of disk close to massive stars, a decline in disk fraction has been observed in regions like NGC 2244, NGC 6611, Pismis 24, NGC 6231, Cygnus OB2, Trumpler 14 and 16 (\citealt{balog2007,guarcello2007,fang2012,damiani2016,reiter2019,gupta2021}).

On the other hand, the phenomenon most commonly associated with the interaction between a star's surface and its encompassing disk is known as magnetospheric accretion \citep{venuti2014}. In this process, the stellar magnetosphere is found to truncate the inner disk extending to a few stellar radii from the stellar surface. Consequently, gas from the inner edge of the disk is channeled onto the photosphere through magnetic field lines, forming accretion columns. When this material impacts the magnetic poles at nearly free-fall velocities, it generates accretion shocks, which manifest through various observable features (see review by \citealt{calvet2000}). These features include excess emission in UV and optical wavelengths above the photospheric flux (\citealt{herczeg2008,ingleby2013,fairlamb2015}), as well as the presence of broad emission lines arising from the heated gas in the accretion columns (\citealt{demarchi2010,rigliaco2012,gangi2022}).

The rate at which the material accumulates onto the host star is termed the mass accretion rate. This is a fundamental property in studies of low-mass star formation since apart from its significance in modeling planet formation theories, it has a direct impact on the final mass of the star and thereby in shaping the form of the initial mass function (IMF) (\citealt{hartmann1998,williams2011}). Observational studies have reported the mass accretion rates in the order of 10$^{-8}$ M$_\odot$/yr at $\sim$1 Myr and is found to decrease with isochronal age by around an order of magnitude at $\sim$10 Myr in accordance with the viscous evolution models. However, there are reports of high accretion rates in older objects at $>$5-10 Myr (\citealt{ingleby2014,venuti2019,manara2020}). 

Historically, assessing accretion rates relied on direct tracers, such as the measurement of UV continuum excess or spectral veiling of photospheric absorption lines, and the intensity of emission lines (\citealt{muzerolle2005,fang2009,lim2014}). However, this approach demanded medium to high-resolution spectroscopy of individual members within a region, limiting its application to nearby regions within the solar neighborhood due to instrument constraints and issues of crowding. To overcome these limitations, \citet{demarchi2010} introduced a method to estimate the accretion properties for a broader sample without the need for spectroscopy by combining the broad band optical photometry with the narrow band H$\alpha$ photometry. Subsequently, this was adopted for several star-forming regions in the Milky Way (\citealt{barentsen2011}) as well in the Magellanic clouds (\citealt{demarchi2017,biazzo2019,carini2022,tsilia2023,vlasblom2023}). Here we use a similar approach to study the disk dynamics of our regions.

The contribution of the environment as well as the interplay between the low-mass star and its surrounding disk through magnetospheric accretion studies have been investigated in several nearby low-mass regions. However such studies on distant massive regions hosting multiple ionising sources are not fully explored owing to observational limitations and high extinction. So, we have analysed two similar but distinct star-forming regions in massive environments beyond the solar neighbourhood (see section ~\ref{sec:intro_w5clusters}). These clusters have moderate core stellar number densities ($\sim$60 pc$^{-2}$ for the East cluster and $\sim$100 pc$^{-2}$ for the West cluster) in between that of low-density regions like Taurus and 25 Ori ($<$10 pc$^{-2}$) (\citealt{esplin2019,suarez2019}) 
and super massive clusters like Arches, RCW 38, NGC 3603 ($>$1000 pc$^{-2}$) (\citealt{beccari2010,stolte2010,muzic2017}). Likewise, our target regions experience feedback from the central massive stars, unlike the low-mass nearby regions that are devoid of O-type stars and lower than that in massive extreme regions that host hundreds of O-type stars. These make them ideal targets to explore the effect of environment on low-mass star formation and disk evolution.  

The paper is structured as follows. In section~\ref{sec:intro_w5clusters} we present the overview of the W5 twin clusters. Section~\ref{sec:datasets} presents the various multi-wavelength datasets used in this study and section~\ref{sec:data_compl} gives the details on the completeness of the photometry. Section~\ref{sec:phy_par} discusses the identification and estimation of the physical parameters of the young stellar objects (YSOs). In section~\ref{sec:analysis} we present the analysis of the properties of the disks such as dependence on stellar parameters and accretion properties. Section~\ref{sec:discussion} discusses the effect of external photoevaporation on the twin clusters and the importance of carrying out this study and section~\ref{sec:conclusion} draws the conclusion of this work.

\section{Twin clusters of W5 - IC 1848 East and West}
\label{sec:intro_w5clusters}

\begin{figure*}
    \centering
  \includegraphics[scale=0.55,trim=30 40 100 60,clip]{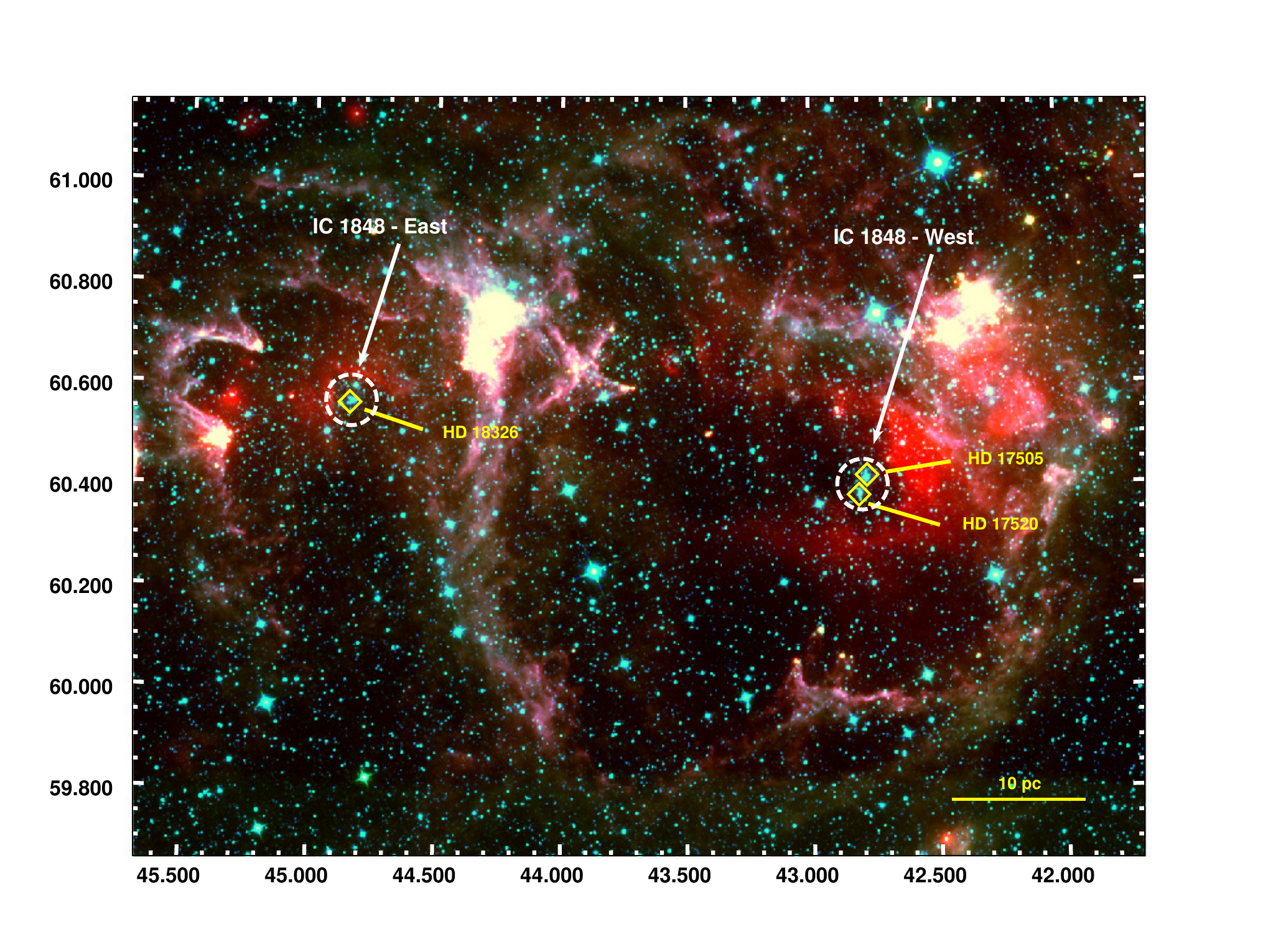}
   \includegraphics[scale=0.32,trim=8 50 50 70,clip]{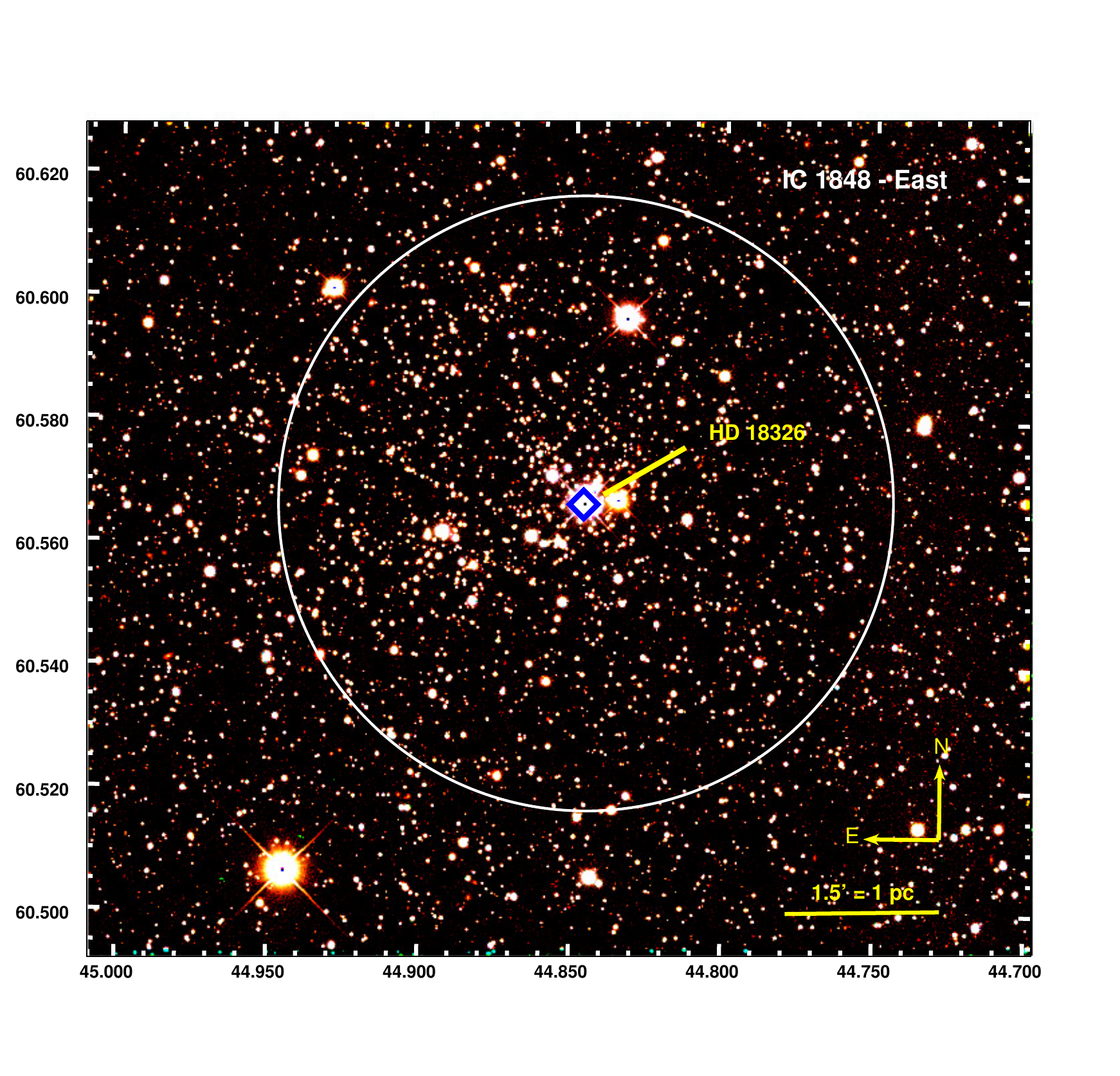}
\includegraphics[scale=0.31,trim=10 30 80 50,clip]{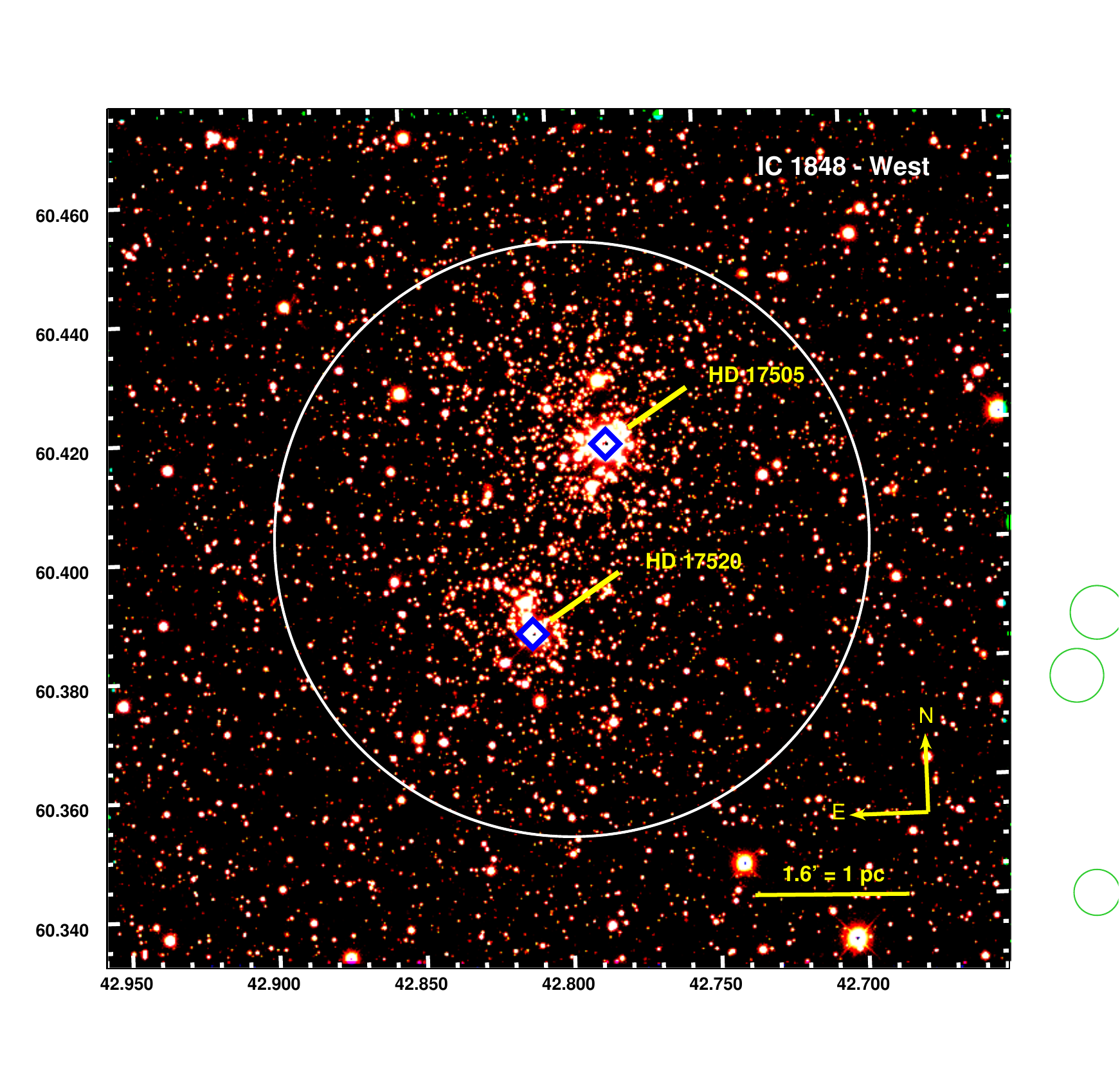}
    \caption{\textit{Top}: WISE color composite image (red-W4, green-W2, blue-W1 band) of the W5 molecular cloud. The massive stars in the twin clusters are marked with yellow diamonds and the dashed circles highlight the areas of the East and West clusters studied here. \textit{Bottom}: NEWFIRM NIR color composite images (red-K, green-H, blue-J band) of the East and West clusters. The white circle shows the area of the cluster considered for the study. The RA and Dec are presented in degrees.}
    \label{fig:w5}
\end{figure*}

In the Cassiopeia OB6 association, W3/W4/W5 collectively known as the heart and soul nebula are well-known giant cloud complexes harbouring young \hii regions of active star formation. W5 or Westerhout 5 has two young clusters namely IC 1848-East and IC 1848-West. Both clusters host multiple O-type stars at their center, which are the main sources of ionisation in the vicinity of the clusters. The clusters are of similar age ($\sim$2 Myr), located at similar distances of $\sim$2 kpc, and due to the presence of massive stars at the center, the surrounding dust is expelled, clearing out a low extinction region of A$_\mathrm{V}$ $\sim$1.5 mag (\citealt{chauhan2011,lim2014,damian2021}). 
On account of these properties, we refer to these clusters as twin clusters. The relatively low extinction towards the clusters enables the observation of the faint low-mass population embedded in them.  The West cluster has a stellar number density ($\sim$40 pc$^{-3}$) almost twice that of the East cluster ($\sim$20 pc$^{-3}$) and UV flux about four times that of the latter \citep{damian2021}. Hence, the W5 twin clusters are diverse in nature in terms of stellar density and UV radiation strength and are ideal sites to understand the role of environment on the star formation and disk evolution processes. Figure~\ref{fig:w5} shows the WISE color composite image of the W5 molecular cloud along with the zoomed-in near-infrared (NIR) color composite image of both clusters. The NIR images are taken from the NOAO archive and observed with NEWFIRM in the 4 m Mayall Telescope (for further details on the images, refer section~\ref{sec:data_newfirm}).

%%%%%%%%%%%%%%%%%%%%%

%%%%%%%%%%%%%%%%%%%%%%

The IC 1848-East cluster is primarily ionized by HD 18326 located at the center of the cluster. HD 18326 is a binary system with two massive stars of spectral types O6.5 V(n)((f)) and O9/B0 V \citep{sota2014}. The center of the cluster is adopted from \citet{damian2021} as RA= 44.8458$^\circ$ and Dec= +60.5667$^\circ$. For the study, we consider sources within a radius of 3$\arcmin$ from the cluster center \citep{damian2021}. The stellar density within the area considered for the study is $\sim$20 pc$^{-3}$ and the total far-UV (FUV) luminosity (L$_\mathrm{FUV}$/L$_\odot$ (ergs/s)) from the central massive stars is $\sim$1.2$\times$10$^{5}$.

IC 1848-West hosts two massive O stellar groups at the center, HD 17505 and HD 17520 both of which are composed of multiple O-type stars. HD 17505 contains at least four O-type stars (O6.5 III((f)), O7.5 V((f)), O7.5 V((f)), O8.5 V) (\citealt{hillwig2006, sota2014, raucq2018}) and HD 17520 is associated with two O-type stars (O8 V and O9:Ve) \citep{sota2014}. The center of the cluster is taken as RA= 42.7958$^\circ$ and Dec= +60.4019$^\circ$ from \citet{damian2021}. The cluster area considered for this study is within 3$^\prime$ radius from the cluster center, limited by the availability of optical data (as detailed in section~\ref{sec:data_cfht}). For the West cluster the stellar density within the area studied is $\sim$40 pc$^{-3}$ and the total FUV luminosity (L$_\mathrm{FUV}$/L$_\odot$ (ergs/s)) from the central O-type stars is $\sim$4$\times$10$^{5}$.

\section{Multiwavelength photometric datasets}
\label{sec:datasets}
For the present study, multiple photometry datasets across a wide wavelength range between $\sim$ 0.5-24 $\mathrm{\mu}$m are used to analyse the W5 clusters. The description of individual datasets is detailed below.

\subsection{Deep near-IR photometry}
\label{sec:data_newfirm}
We extracted deep NIR photometry for the twin clusters from observations with the wide-field IR imager NEWFIRM (NOAO Extremely Wide Field Infrared Imager; \citealt{probst2004}) in the 4 m Mayall Telescope at Kitt Peak National Observatory, Arizona (PI: Guy Stringfellow). NEWFIRM has a field of view of 28$^\prime$ $\times$ 28$^\prime$ with a pixel scale of 0.4$^{\prime\prime}$. We obtained the calibrated, stacked, and mosaiced images of the twin clusters in J, H, and K$_s$ bands from the NOAO (National Optical Astronomical Observatory) archive\footnote{\url{http://archive1.dm.noao.edu/search/query/}}. The photometry in all three bands was performed with IRAF. Using the DAOFIND task we identified the point sources in K$_s$ band with signal 5$\sigma$ above the background to avoid any false detection or artifacts in the image. We then conducted a visual inspection of the detected sources and removed spurious detections mainly around the cluster area in the field. This step is crucial to avoid contamination of the sample with artificial detections, and the number of spurious sources removed was minimal ($<$0.5\%), making them inconsequential to the member statistics. The refined source list was then used to extract sources in the J and H bands as well. We performed psf photometry of these sources using the ALLSTAR routine of IRAF (e.g. \citealt{jessy2012,jessy2016}). For absolute photometric calibration, we used the common sources in our catalog and the Two Micron All Sky Survey (2MASS) catalog \citep{cutri2003} with the quality flag `A' in all the three bands and a match radius of 1.0$^{\prime\prime}$. The zero point correction term with respect to 2MASS photometry has been applied to our NEWFIRM photometry for individual bands to calibrate it. For sources with J $<$ 13.5 mag, the photometry in all three filters is likely to be saturated. Hence for those sources, we replace the J, H, and K$_s$ magnitudes with 2MASS photometry \citep{cutri2003}.

\subsection{Optical photometry}
\subsubsection{CFHT Megaprime}
\label{sec:data_cfht}
We have used the optical photometry for the IC 1848-West cluster from observations with the wide-field optical imager Megaprime in the 3.6 m Canada-France-Hawaii Telescope (CFHT; \citealt{puget2004}). The field of view of Megaprime is 0.96$^\circ$ $\times$ 0.94$^\circ$ with a pixel scale of 0.187$^{\prime\prime}$/pixel. The observations were taken in August 2004 (PI: Jean-Charles Cuillandre) and we downloaded the calibrated images from the CADC (Canadian Astronomy Data Center) archive\footnote{\url{https://www.cadc-ccda.hia-iha.nrc-cnrc.gc.ca/en/search/}} in g and r bands. 

The source extraction was performed using IRAF. We aligned and average combined all the frames in both the filters separately and performed psf photometry to extract the sources in r-band and used the same source list to extract the point sources in g-band. The radius of the cluster considered for further analysis is limited by the coverage in both these bands. We consider the cluster area within 3$^\prime$ from the cluster center which is the region covered uniformly in all the exposures in both g and r filters. The zero point correction was estimated using the common sources in the CFHT catalog and the Pan-STARRS survey (PS1) catalog \citep{chambers2016} with error in g and r bands $<$0.1 mag and the correction was applied to the CFHT photometry to calibrate the sources. There was no color correction required for this calibration.

\subsubsection{DOT ADFOSC}
\label{sec:data_dot}
For the IC 1848-East cluster we use the optical photometry obtained using ADFOSC (ARIES Devasthal Faint Object Spectrograph Camera) on the 3.6 m Devasthal Optical Telescope (DOT). ADFOSC has a field of view of 13.6$^\prime$ $\times$ 13.6$^\prime$ with a pixel scale of 0.2$^{\prime\prime}$. We observed the East cluster with ADFOSC in g and r bands during December 2020. Multiple 60s exposures were taken in both the bands (32 and 21 exposures in g and r bands, respectively). Using the various packages in IRAF, we bias subtracted, flat field corrected, aligned, and stacked all the frames for both filters. We used the DAOFIND task to identify the point sources in r-band and used the PSF and ALLSTAR routines to perform psf photometry. This source list was used to extract the sources in g-band as well. To get the wcs information of our sources we use the 'ccmap' and 'mscred' tasks in IRAF. The photometric calibration was performed following a similar procedure as the West cluster.

\subsubsection{IGAPS}
\label{sec:data_igaps}
We obtain the INT Galactic Plane Survey (IGAPS) photometry for both the W5 clusters. IGAPS comprises of the optical photometric surveys, IPHAS (INT Photometric H$\alpha$ Survey of the Northern Galactic Plane, \citealt{drew2005}) and UVEX (UV-excess Survey of the Northern Galactic Plane, \citealt{groot2009}), observed with the Wide Field Camera (WFC) on the 2.5 m Isaac Newton Telescope (INT). The WFC has a pixel size of 0.33$^{\prime\prime}$/pixel with a field of view $\sim$0.22 deg$^2$. We use the data obtained with r, i, and H$\alpha$ filters for an area of 3$^\prime$ radius from the center of both clusters.

\subsubsection{Pan-STARRS}
\label{sec:data_panstarrs}
We use the optical data from the Pan-STARRS survey DR1 (PS1, \citealt{chambers2016}) in g, r, i, z, and y filters for both the W5 clusters. We obtain the photometry for the cluster area as described in section~\ref{sec:intro_w5clusters}.

\subsection{Mid-IR photometry}
\label{sec:data_spitzer}

{\it Spitzer}-IRAC images in  3.6, 4.5, 5.8, and 8.0 $\mu$m bands are obtained from the {\it Spitzer} archive\footnote{\url{https://irsa.ipac.caltech.edu/applications/Spitzer/SHA/}}.  These observations were taken as part of the deep imaging survey of the dense young clusters surrounding the massive O-type stars in the interior of the W5 star-forming complex (PI: G. Fazio, Program ID: 40448). The processing of the images and extraction of photometry have been performed similarly to the steps given in \citet{jessy2017}. Since these observations were one of the deepest by {\it Spitzer}, we obtain the photometry of fairly low-mass objects (see section~\ref{sec:data_compl} for details).  MIPS 24 $\mu$m photometry has been taken from \citet{koenig2008}.

\subsection{Compilation of multiband catalogs and identification of disk sources}
\label{sec:multiwave_catalog}
We prepared the multiwavelength catalog by combining the photometry from the respective filters as discussed above for both clusters within 3$^\prime$ radius from their center. For the IC 1848-East cluster, we crossmatch the sources in the NEWFIRM dataset with counterparts in at least one of the filters of other datasets such as {\it Spitzer}, DOT ADFOSC, Pan-STARRS, and IGAPS photometry with a match radius 1$^{\prime\prime}$. The combined catalog consists of 2226 sources and the photometric completeness in individual bands is given in section~\ref{sec:data_compl}. Similarly, for the sources in the IC 1848-West cluster, we combined the NEWFIRM dataset with {\it Spitzer}, CFHT Megaprime, Pan-STARRS and IGAPS photometry with a match radius of 1$^{\prime\prime}$. Here the combined catalog consists of 3342 sources with their completeness in each band presented in the following section. In both catalogs, we use the photometry of the sources with uncertainty $<$0.2 mag in the NIR and MIR filters which are the data primarily used in our analysis.

From the list of sources for both clusters, we identify the YSOs with IR excess based on the classification scheme given by \citet{gutermuth2009}. This method identifies YSOs based on several criteria using the NIR photometry in $J,H,$ and $K_s$-bands  and Spitzer-IRAC photometry. The first step was to eliminate the contaminants such as PAH emitting galaxies, shock excited emissions, PAH contaminated apertures, and broad-line AGNs. From the remaining sources,  we followed the three-phase process outlined in \citet{gutermuth2009}, to classify the IR excess sources into Class I and  Class II category. The remaining sources were termed non-excess sources, which could be either members or non-members of the clusters. Among the sources in the East cluster, 327 show excess emission due to the presence of disks (among which 6 are classified as Class I and the remaining are Class II), and 1899 sources are identified as non-excess sources. Whereas for the West cluster, we have 342  excess sources (all of them identified as Class II) and 3000 non-excess sources.  Since the number of Class I sources is very minimal, we consider the Class I and Class II sources together as IR-excess or disk sources hereafter.  The membership analysis of the non-excess sources is performed in section~\ref{sec:sed_vosa}.

\section{Data completeness}
\label{sec:data_compl}
Data incompleteness is a common issue with photometric data, especially towards the fainter end, which in general can be attributed to various factors such as variable extinction in the surveyed region, sensitivity of different observations, and crowding. To ensure the accuracy of the analysis, it is important to assess the photometry completeness. 

We estimate the completeness in different bands with respect to the NEWFIRM H band which has the highest number of source detections among all bands. Figure~\ref{fig:completeness} shows the completeness levels in various bands, including NEWFIRM J and K$_s$, as well as the Spitzer IRAC1 and IRAC2 bands, all relative to the NEWFIRM H band for both studied clusters.  The 90\% completeness in each band with respect to H band magnitude is marked by the turnover point in the histogram (\citealt{maia2016,damian2021,adamian2023apj}). Notably, the H band photometry for both clusters attains 90\% completeness up to 18 mag which corresponds to $\sim$0.07 M$_\odot$ (by incorporating the respective distance and extinction towards each cluster derived in section~\ref{sec:dist_ext} and using \citet{baraffe2015} 2 Myr isochrone).

To gauge the completeness of other filters in our dataset, we count the number of counterparts to the H band detections within each 1 mag bin up to H=18 mag. This approach provides insights into the relative completeness across various bands. In Table~\ref{tab:completeness}, we present a comprehensive list of all the filters used in our study for both clusters, along with their respective completeness metrics relative to the H band. Our multiwavelength dataset exhibits completeness levels exceeding 80\% across a minimum of 5 bands in both clusters (mainly in NEWFIRM J, H, K$_s$ and Spitzer IRAC1 and IRAC2), underlining the robustness and comprehensiveness of our data compilation. Notably, with the deep imaging observations of the complex with {\it Spitzer}, the IRAC1 and IRAC2 photometry is deeper by $\sim$ 2 mag (see figure~\ref{fig:completeness}) compared to typical observations of {\it Spitzer} (eg. \citealt{jessy2013}).    Most of the other bands show $>$50\% completeness in both the clusters namely in Spitzer - IRAC3, PanSTARRS - r, i, z, y, and IGAPS i band. Additionally, we have $>$50\% completeness for the West cluster in Megaprime - g, r, whereas for the East cluster in ADFOSC - g, r, and IGAPS H$\alpha$, r. It is worth noting that the completeness of our datasets may be affected, particularly in regions closer to the cluster's center, owing to the presence of bright and massive stars that can affect the source detection. We measure an area of $\sim$5$\arcsec$ to 6$\arcsec$ radius around the central bright source that maybe affected due to saturation. Additionally, at the H-band completeness limit of 18 mag, the innermost 30$\arcsec$ region and the 30$\arcsec$ to 1$\arcmin$ annular region around the cluster center have a number density that is comparable to about $\sim$85-90\% of the number of sources detected in the entire cluster area at this magnitude.

\begin{table}
    \centering
    \begin{tabular}{l|l|l}
        \hline
        Filters & IC 1848 - West & IC 1848 - East\\
        \hline 
        NEWFIRM H & 2244 & 1541\\
        NEWFIRM J & 2210 (98\%) & 1532 (99\%)\\
        NEWFIRM K$_s$ & 2236 (100\%) & 1538 (100\%)\\
        Spitzer IRAC1 & 1895 (84\%) & 1436 (93\%)\\
        Spitzer IRAC2 & 1894 (84\%) & 1426 (93\%)\\
        Spitzer IRAC3 & 1273 (57\%) & 1108 (72\%)\\
        Spitzer IRAC4 & 926 (41\%) & 685 (44\%)\\
        Megaprime g & 1301 (58\%) & -\\
        Megaprime r & 1544 (69\%) & -\\
        ADFOSC g & - & 1057 (69\%)\\
        ADFOSC r & - & 1125 (73\%)\\
        PanSTARRS g & 659 (29\%) & 483 (31\%)\\
        PanSTARRS r & 1194 (53\%) & 926 (60\%)\\
        PanSTARRS i & 1519 (68\%) & 1167 (76\%)\\
        PanSTARRS z & 1749 (78\%) & 1296 (84\%)\\
        PanSTARRS y & 1605 (72\%) & 1216 (79\%)\\
        IGAPS i & 1310 (58\%) & 924 (60\%)\\
        IGAPS H$\alpha$ & 817 (36\%) & 945 (61\%)\\
        IGAPS r & 964 (43\%) & 812 (53\%)\\
        \hline
    \end{tabular}
    \caption{Data completeness in all the filters with respect to NEWFIRM H band upto H=18mag.}
    \label{tab:completeness}
\end{table}

\section{Physical parameters of the clusters}
\label{sec:phy_par}
\subsection{Distance and extinction}
\label{sec:dist_ext}
The distance and extinction of the W5 clusters are adopted from \citet{damian2021}. Here we briefly discuss the estimation of these two parameters (for details refer \citealt{damian2021}). The distance to the individual pre-main sequence (PMS) members within the cluster area is obtained from \citet{bailerjones2018} with a parallax uncertainty $\leq$0.2 mas. To derive the mean distance to the cluster, a Gaussian curve is modeled over the distance distribution of all the PMS members. This distribution is further converged to sources within 1$\sigma$ deviation from the peak and the mean and standard deviation of this iterated Gaussian curve are considered as the average distance and corresponding uncertainty. The distance to the East cluster is taken as 2380$\pm$510 pc and the West cluster as 2220$\pm$420 pc. Similar distances have been estimated for the W5 clusters in previous studies. \citet{lim2014} derived the distance to the West cluster as 2.2 kpc.

In a similar manner, the candidate PMS members are used to estimate the average extinction towards the cluster.  We adopt from \citet{pecaut2013} the average (J-H) intrinsic colour of K and M stars listed in their table to estimate the E(J-H) colour excess. The K band extinction for each of the PMS sources is derived by adopting the extinction ratio from \citet{cardelli1989}. A Gaussian curve is fit to the histogram distribution of the extinction of individual sources in both the clusters (e.g. \citealt{swagat2023}). The mean and standard deviation of the Gaussian fit are taken as the average extinction towards the cluster and the uncertainty, respectively. The K-band extinction (A$_\mathrm{K}$) of the East cluster is 0.21$\pm$0.07 mag and the West cluster is 0.14$\pm$0.05 mag. The small deviation of around 0.05-0.07 mag suggests that the level of reddening variation in the clusters is relatively low within the selected radius. Due to the relatively low and uniform extinction within the cluster area, we are able to probe the very low-mass objects in the region (as can be seen in figure~\ref{fig:hrd}) with completeness above 80\% in at least five bands down to $\sim$0.07 M$_\odot$. We underline that both clusters are situated at nearly the same distance in the molecular cloud and experience comparable levels of extinction.

\subsection{Age and Mass estimation}
\subsubsection{Spectral energy distribution using VOSA}
\label{sec:sed_vosa}
The age and mass of the cluster members are among the most fundamentally important physical parameters to understand the evolution of the system. The conventional approach to determine these attributes involves employing the Hertzsprung-Russell diagram (HR diagram). In this method, the positions of stars are compared with the theoretical PMS evolutionary tracks and isochrones. Here we use the VO SED Analyser (VOSA)\footnote{\url{http://svo2.cab.inta-csic.es/theory/vosa/}}, an online tool that generates a SED based estimation of various parameters by comparing the observed photometry with theoretical models (\citealt{damian2021,guzman2021,gupta2024}). 

We make use of the complete multiband photometry as discussed in section~\ref{sec:multiwave_catalog} for the non-excess sources in both clusters but limit the usage to the optical and NIR data, up to K-band, for disk sources that emit excess at longer wavelengths. The respective distance and extinction of both clusters are given as initial parameters with metallicity at solar metallicity. After correcting for the distance and extinction, VOSA analyzes the observed spectral energy distribution (SED) by comparing it to synthetic photometry generated from the BT-Settl-CIFIST theoretical models \citep{baraffe2015}. By utilizing the reduced chi-square minimization technique, we acquire the most suitable model that fits the data and yields associated physical parameters, including luminosity, effective temperature, and radius for each source. With the derived luminosity and effective temperature, the age and mass are estimated based on their position to the nearest isochrones and evolutionary tracks from \citet{baraffe2015}. 

\begin{figure*}
    \centering
    \includegraphics[width=\columnwidth]{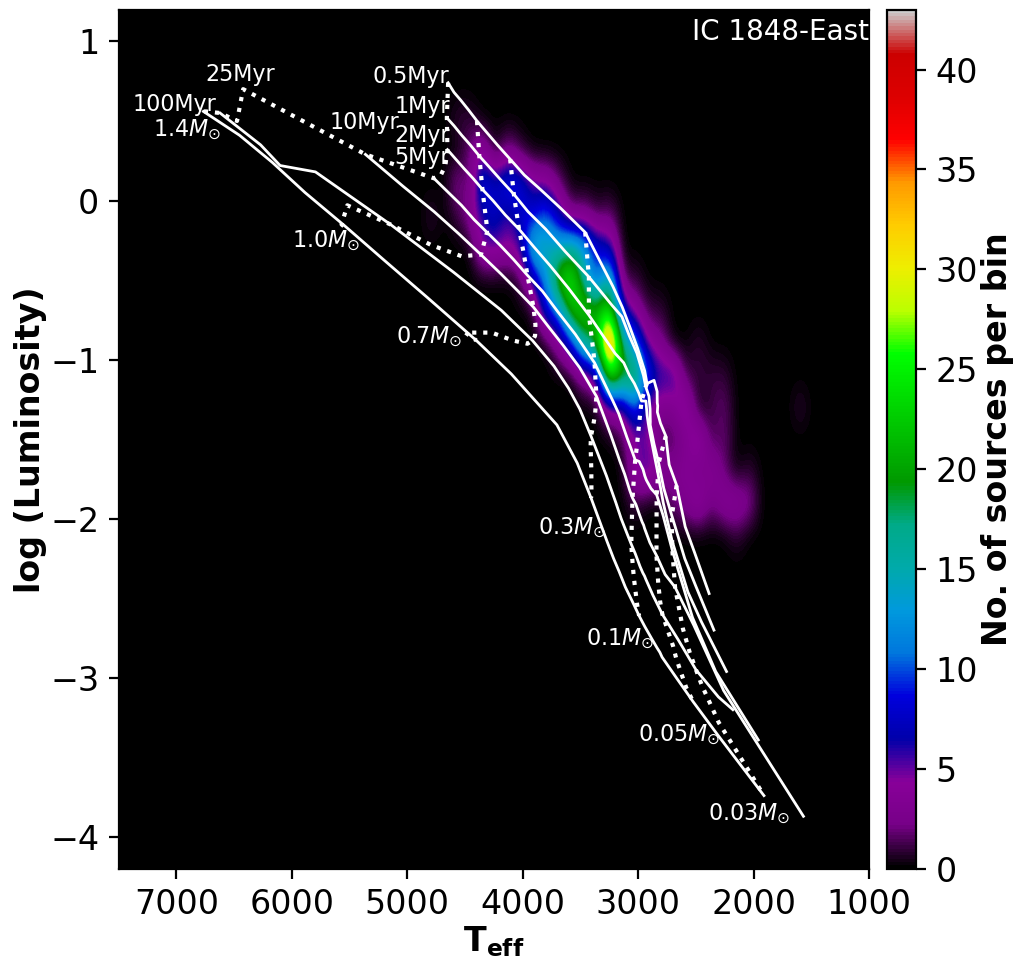}
    \includegraphics[width=\columnwidth]{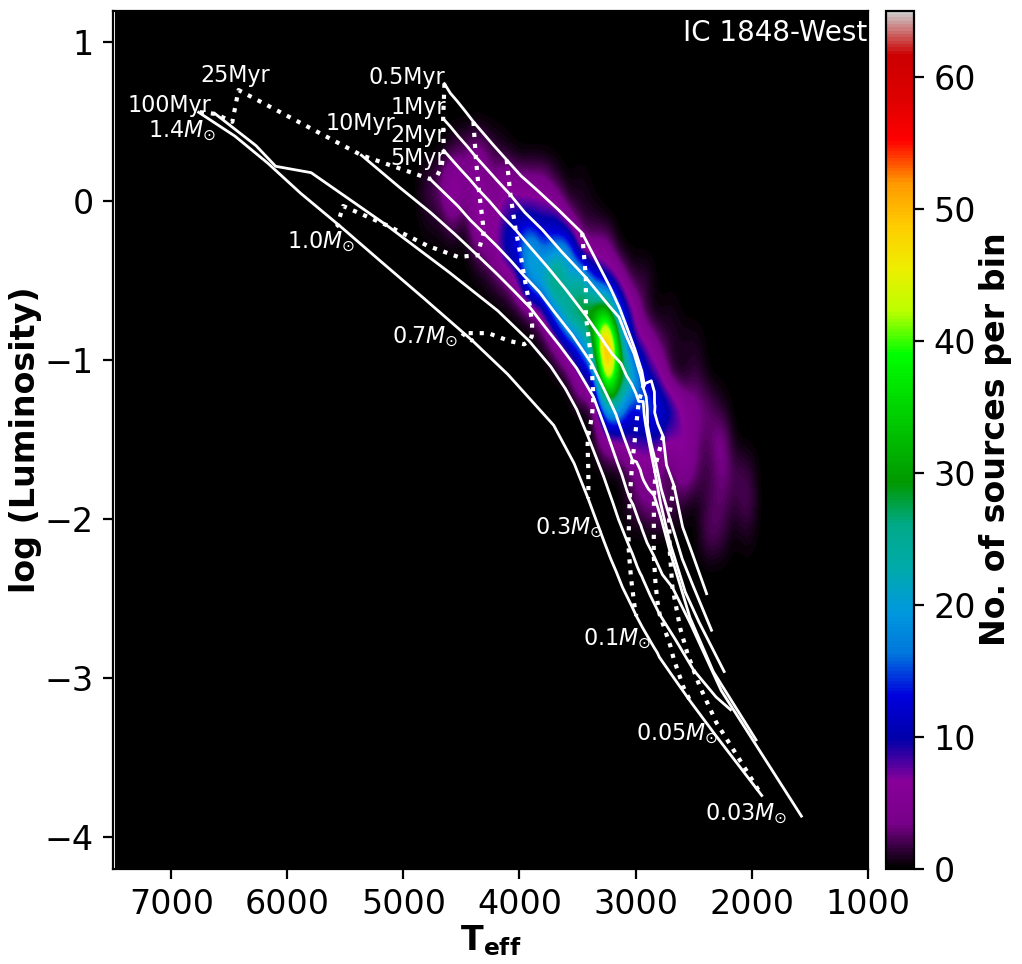}
    \caption{Hess plot of HR diagram of the PMS sources with age below 10 Myr in the East (left) and West (right) clusters. The dotted and continuous lines are the evolutionary tracks and isochrones from \citet{baraffe2015} for various ages and masses.}
    \label{fig:hrd}
\end{figure*}

In Figure~\ref{fig:hrd} we show the HR diagram as a Hess plot for both the clusters where the luminosity and effective temperature are estimated using VOSA. As we are interested in the PMS members of the twin clusters, we consider sources with age $<$10 Myr. For the sources lying above the 0.5 Myr isochrone, we estimate the mass by extrapolating the \citet{baraffe2015} evolutionary tracks and we approximate the age to be $<$0.5 Myr (\citealt{almendros2023,adamian2023apj}).  With the multiband SED analysis, we determined the age and mass of 824 and 1174 sources with age $<$10 Myr in the East and West clusters respectively. In the following sections, we consider these sources with age $<$10 Myr as the PMS cluster members. 

The average age of the PMS cluster members is determined by calculating the median age of the sources ranging from 0.5 to 10 Myr. The median age and corresponding median absolute deviation of the East and West clusters are 2$\pm$1 and 2.1$\pm$1 Myr. The dependence on stellar evolutionary models for the determination of cluster age and its associated uncertainties is a known problem (\citealt{bell2013,soderblom2014}). The following analysis and inferences from the results on the evolution of disks in high radiation environments are derived assuming a young age for the clusters, which is subjective of the \citet{baraffe2015} models used here.

The completeness of this analysis is limited by the mass range of the evolutionary tracks of the \citet{baraffe2015} model. Hence the mass of the PMS members identified based on this analysis ranges between 0.01-1.4 M$_\odot$. We see that though the number of PMS sources identified in both the clusters varies, they have similar distributions in the HR diagrams given in Figure~\ref{fig:hrd},  with the peak density between 0.2-0.3 M$_\odot$, highlighting one of the twinning characteristics of the clusters. This is consistent with the characteristic peak mass of the IMF derived in \citet{damian2021}. These two HR diagrams presented in Figure~\ref{fig:hrd} are some of the deepest HR diagrams for distant massive star-forming regions probing down to the substellar regime.

\section{Analysis and Results}
\label{sec:analysis}
With the identified YSOs in both regions, we analyse the various properties of the disks such as the dependence of disk fraction on stellar parameters like age and mass as well as the accretion properties of the disks and their correlation with stellar mass and age. We study the effect of environment due to the presence of massive stars in the region on the evolution of disks around the low-mass stars.

\subsection{Distribution of disk and diskless YSOs in CMD}
\label{sec:yso_cmd}
In order to validate the nature of the PMS sources identified through the SED analysis, we trace their distribution in various color-magnitude diagrams (CMDs). Circumstellar disks surrounding young stars in the PMS phase experience heating from the central star. As a result, an inner wall forms within the disk, which absorbs and re-emits energy. This emission leads to noticeable excesses observed predominantly in MIR regime beginning from K-band to longer wavelengths in comparison to stars in the main sequence. Consequently, the presence of IR excesses in PMS stars serves as strong evidence for the existence of circumstellar disks (\citealt{hillenbrand1998,yao2018}). However, it is important to note that the absence of excess does not definitively indicate the absence of a circumstellar disk, as it could be attributed to an inner hole within the disk or due to the disk orientation.

\begin{figure*}
    \centering
    \includegraphics[width=\textwidth]
    {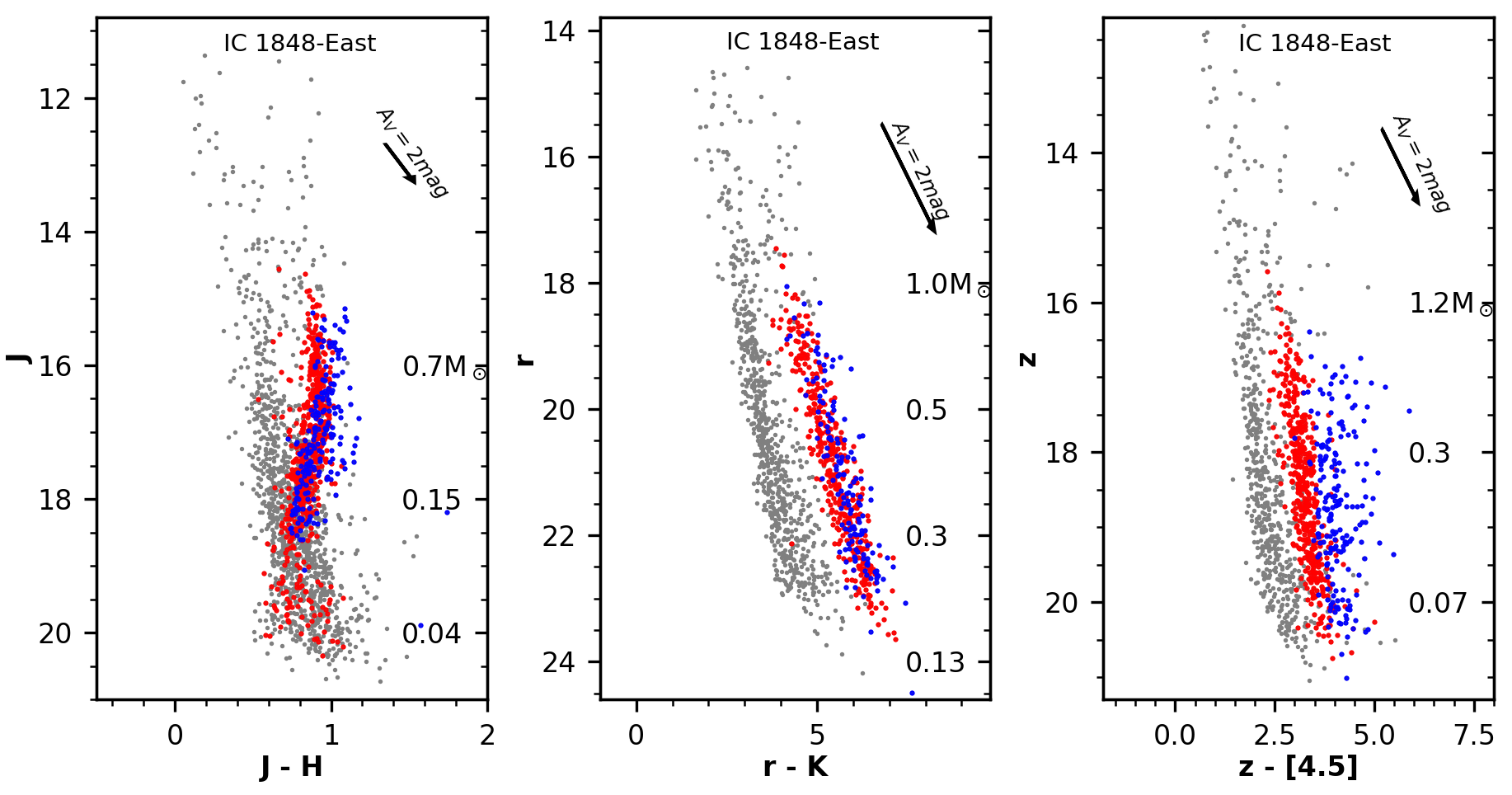}
    \includegraphics[width=\textwidth]
    {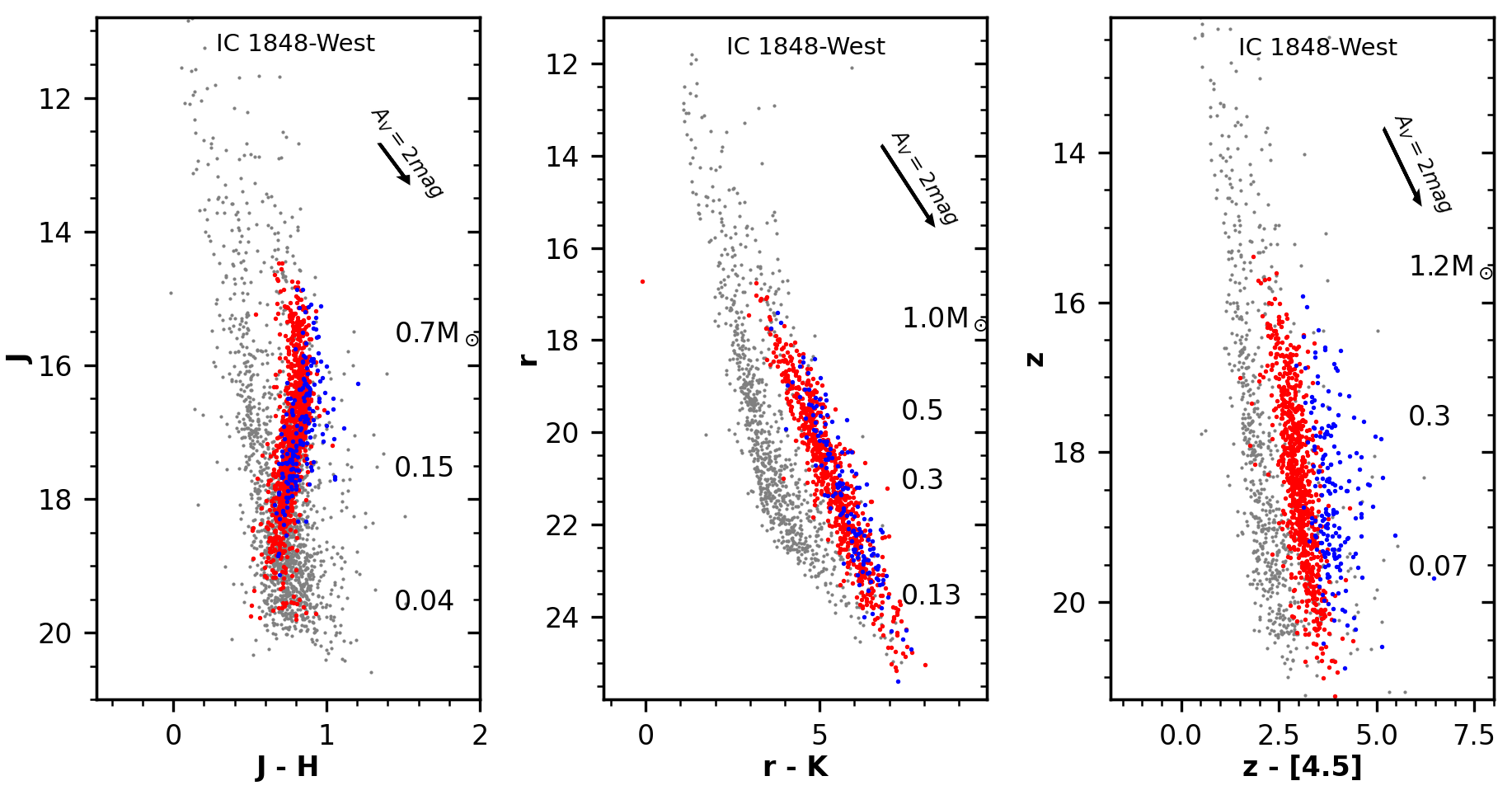}
    \caption{CMDs of the sources within the cluster area marked in grey. The candidate members identified from the SED analysis are marked as red and blue dots, where, the former are the non-excess and the latter are the excess sources of age $<$10 Myr. \textit{Top}: CMDs of IC 1848-East with NEWFIRM J, H, and K$_s$ bands, ADFOSC r band, Pan-STARRS z band and Spitzer IRAC2 filter. \textit{Bottom}: CMDs of IC 1848-West with NEWFIRM J, H, and K$_s$ bands, Megaprime r band, Pan-STARRS z band, and Spitzer IRAC2 filter.}
    \label{fig:cmds}
\end{figure*}

Among the 824 PMS sources in the East cluster identified in section~\ref{sec:sed_vosa} based on SED analysis, 203 sources are disk-bearing YSOs and the remaining 621 sources are non-excess sources. Similarly, in the West cluster among the 1174 PMS sources, 184 are excess emitting sources and 990 are non-excess sources. Figure~\ref{fig:cmds} shows various CMDs using photometry from different filters for both clusters. We highlight the young, excess, and non-excess sources that trace a distinct PMS branch to the right side of the CMDs. The well-defined shape of the sequence indicates the uniform extinction within both clusters. The excess sources are scattered rightward of the non-excess sources at longer wavelengths. The distribution of the PMS sources in all the CMDs conforms the results of the independent SED based membership analysis.  

The following analysis to investigate the disk characteristics around low-mass stars is performed based on the 824 and 1174 PMS sources in the East and West clusters, respectively. We note that the astrometric solutions from Gaia DR3 were not utilised for any membership analysis due to the relatively large distances of these regions, which limit the detection of these PMS sources. Additionally, only $<$10\% of these PMS cluster members have reliable astrometric data with parallax uncertainty $<$0.2 mas.

%%%%%%%%%%%%%%%%%%%%%%
%%%%%%%%%%%%%%%%%%%%%

\subsection{Properties of protoplanetary disks around YSOs}
As discussed in section~\ref{sec:intro}, the evolutionary timescale of disks sets an important constraint on planet formation. The circumstellar disks provide raw materials for planet formation, hence disk dissipation and its dependence on external factors like the mass of the host star \citep{ribas2015}, metallicity of the star-forming region (\citealt{yasui2009,ercolano2010,ashraf2023,gehrig2023,patra2024}), stellar density \citep{zwart2016} and external photoevaporation \citep{haworth2018,haworth2023,parker2021,winter2022} play a deciding role in when and where planets can form (refer review by \citealt{williams2011}). This dependence is generally quantified by the disk fraction which describes the percentage of PMS sources that host a disk around them (i.e. the ratio of the disk sources among the cluster members to the total number of cluster members in each region). Although the disk fraction is found to decrease as a function of age, it is generally observed that in young clusters in the solar neighbourhood, below $\sim$5 Myr, the disk fraction is $\sim$40-80\% (\citealt{haisch2001,samal2012,michel2021,bdamian2023jaa,roccatagliata2023,patra2024}).  In this study, since the excess and non-excess sources are identified primarily based on the NIR and MIR bands for which we have above 80\% completeness at H=18 mag, our data is complete down to $\sim$0.07 M$_\odot$. Here to quantify the disk fraction we apply the mass completeness limit approximated to 0.1 M$_\odot$ (refer section~\ref{sec:data_compl}) and use sources with mass above this limit in both clusters. For the East cluster, the estimated disk fraction and the corresponding Poisson error taken as the uncertainty, is $\sim$27$\pm$2\% and for the West cluster, it is $\sim$17$\pm$1\%. 

The difference in the disk fraction between the two clusters is not likely due to the evolutionary gradient as suggested by similar previous studies of other star forming regions. \citet{allen2012} examined the two sub-clusters in the Cep OB3b association and found an 18\% difference in disk fraction between them, attributed to the difference in their ages of 1 Myr. But for the twin clusters of W5, as can be seen from Figure~\ref{fig:hrd} as well as the estimated median age of the clusters, the temporal difference may not be a contributing factor to the observed variation in the disk occurrence. Given that these are young clusters ($\sim$2 Myr), the estimated disk fraction is below the expected range as seen in other nearby regions (\citealt{ribas2014,richert2018,bdamian2023jaa}). We discuss the possible causes for this decrease in the sections below, where we analyse the influence of the host stellar mass as well as the external radiation on the evolution of protoplanetary disks around the cluster members.

\subsubsection{Dependence of disk fraction on stellar mass}
\label{sec:df_mass} 

\begin{figure}
    \centering
    \includegraphics[width=\columnwidth]{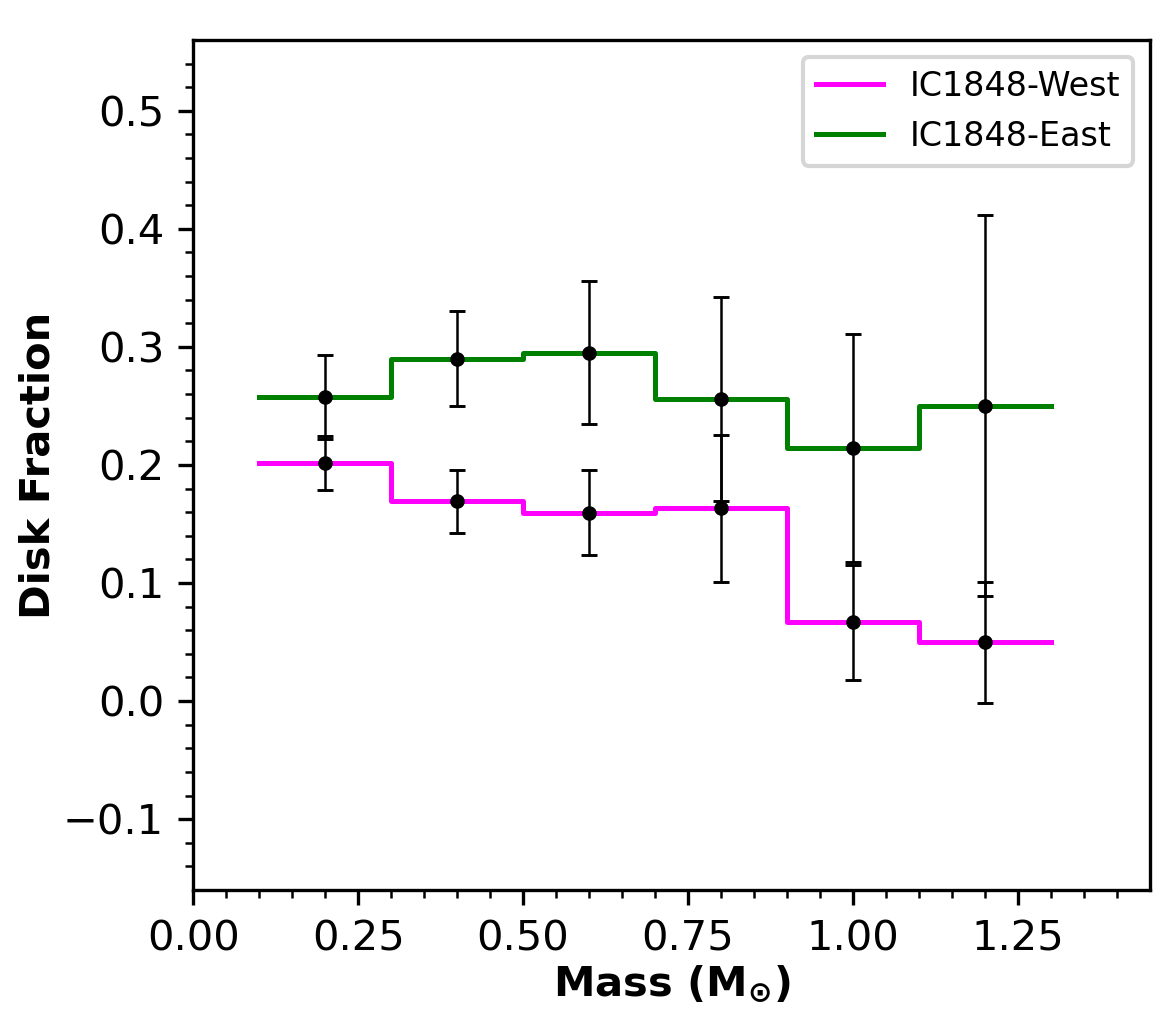}
    \caption{Disk fraction vs stellar mass. The disk fraction measures the number of disk bearing sources to the total number of stars in every 0.2 mass bin. The error bar denotes the Poisson error at each bin.}
    \label{fig:df_mass}
\end{figure}

With the derived disk fraction we investigate the link between the frequency of disks around the PMS sources and their host mass. In figure~\ref{fig:df_mass} we show the distribution of disk fraction in both the regions as a function of mass where the sources are binned every 0.2 dex in mass. In this figure, we consider sources above the mass completeness limit of 0.1 M$_\odot$ in both regions. The overall disk fraction of the West cluster across the entire mass range considered here is lower than that measured for the East cluster. We observe a slight decline in the abundance of excess emitting sources as the stellar mass increases among the members of the West cluster than the East cluster. This diminished occurrence of disks in higher mass stars might indicate an accelerated evolution of disks (\citealt{luhman2022,pfalzner2022}). Conversely, the relatively higher disk fraction among low-mass stars could be ascribed to a longer disk lifetime. 

The shorter disk dispersal timescale of high-mass stars can be due to the higher efficiency in internal photoevaporation or accretion onto the host star. Previous studies have shown that the disk dispersal is dependent on the mass of the host star (\citealt{carpenter2006,roccatagliata2011,ribas2015}). \citet{kennedy2009} reported this dependence by examining nine young clusters ($\sim$1-10 Myr) whose intermediate mass stars were found to lose their disks earlier than solar mass stars. \citet{wilhelm2022} analysed the dependence of disk lifetime on the mass of the host star and found that for stars below 0.5 M$_{\odot}$, external photoevaporation dominates, whereas for higher mass stars, both internal photoevaporation and accretion dominate. A similar correlation was observed by \citet{juan2015} who studied the low-mass stars and brown dwarfs in the 25 Orionis group and found that the disk dissipation efficiency inversely correlates with the mass of the central object.

It has been observed that the lifetime of disks around K-type PMS stars can last up to $\sim$5 Myr \citep{pecaut2016} whereas those around M-type stars last longer with fractions as high as 9\% at 20 Myr \citep{silverberg2020}. With a sample of nearby clusters ($<$200 pc), \citet{pfalzner2022} deduce the disk lifetime of low-mass stars to be 5-10 Myr, whereas for high-mass stars to be 4-5 Myr. This prolonged lifetime of disks around low-mass stars increases the timescale available for the formation of planets. However, in the case of the W5 twin clusters, we observe that the overall disk frequency is low at $\sim$20\% for an average age of $\sim$2 Myr which suggests that the disks have undergone rapid dissipation within the clusters (see sections \ref{sec:df_flux} and \ref{sec:effects_radiation} for details).

%%%%%%%%%%%%%%%%%%%%%%%%%%%%%%%
%%%%%%%%%%%%%%%%%%%%%%%%%%%%%%%%

\subsubsection{Dependence of disk fraction on external radiation}
\label{sec:df_flux}
In the clustered mode of star formation, most low-mass stars spend their early years of formation in regions of enhanced stellar density. Hence the environment influences their evolution, mainly in the form of photoevaporation due to strong radiation fields and tidal truncation due to close encounters (\citealt{clarke1993,alexander2014,facchini2016,andrews2020}). \citet{winter2018} compared these two effects in varying cluster environments and found that external photoevaporation always has a dominant effect over tidal truncation. Theoretical studies also report that the effect of encounters can strip the outer regions of the disk thereby influencing the disk radii more than other disk properties such as disk mass \citep{rosotti2014}. Here, we look at the environmental effects on the disk in terms of the strength of the ionising radiation from the nearby massive stars. The FUV radiation from the massive stars is observed to photodissociate the disk in its vicinity driving strong disk mass loss rates \citep{haworth2023}. This affects the disk structure (\citealt{ansdell2017,eisner2018}) and its lifetime (\citealt{guarcello2016,terwisga2020}).

\begin{figure}
  \centering
    \includegraphics[width=\columnwidth]{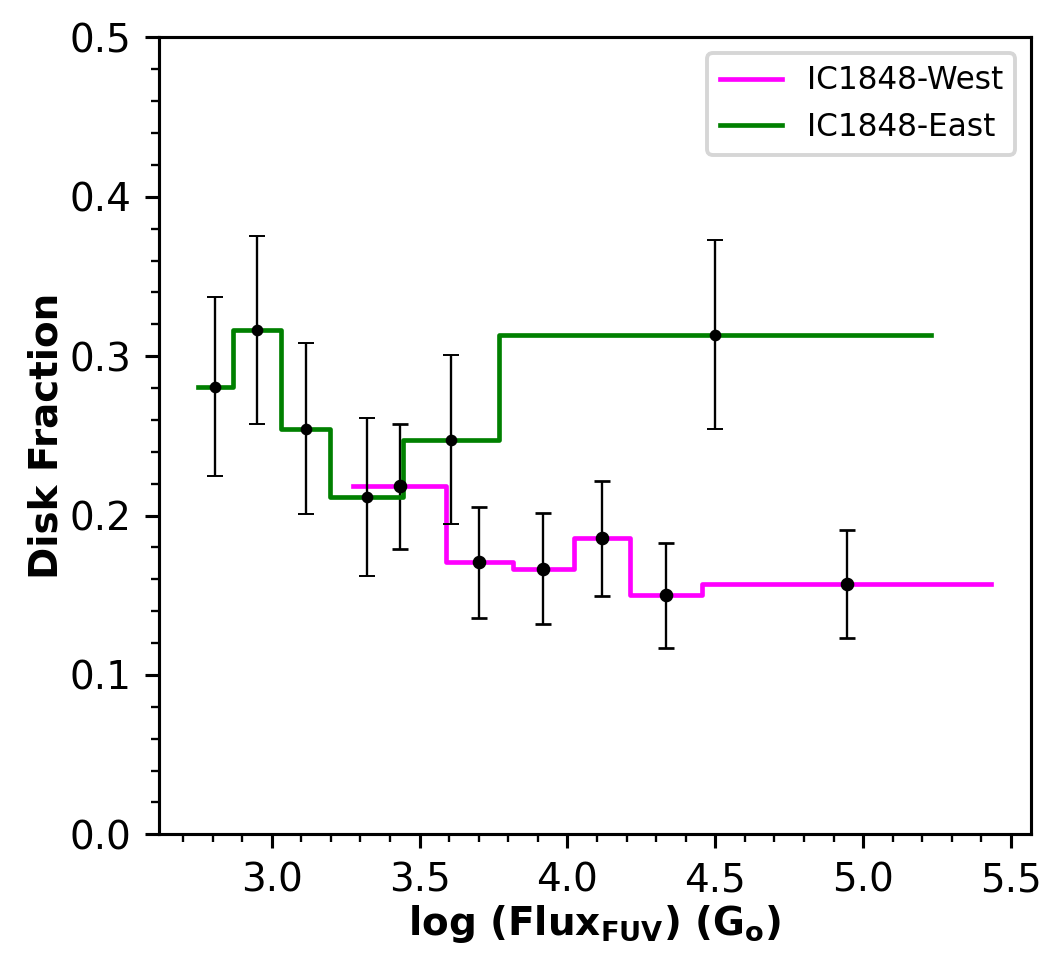}
    \caption{Disk fraction vs FUV flux distribution. The disk fraction is measured in bins of varying size where the number of diskless sources is uniform across all bins and the error bars denote the Poisson error.}
    \label{fig:df_flux}
\end{figure}

In order to understand the effect of massive O-type stars present in the cluster on the evolution of disk, we study how the disk fraction varies as a function of the FUV radiation.  To do so, we estimate the FUV flux emitted by each O-type star in the cluster and the incident on the PMS population associated with them. We note that the twin clusters host only a few B-type stars (2 stars in the West cluster and 4 stars in the East cluster) as listed in the SIMBAD database within the area considered for the study. However, all of them are late B-type stars, later than B5, radiating UV luminosity log(L$_\mathrm{FUV}$/L$_\odot$) below 2.5 \citep{thompson1984} which is $\sim$2 to 3 orders of magnitude lower than the radiation emitted by the O-type stars in the region. Additionally, similar studies in literature (e.g. \citealt{guarcello2007,stolte2010,guarcello2016,mauco2023}) that assess the environmental effect on disk evolution have quantified the integrated flux mainly from the O-type stars and in some cases from massive B-type stars as well. Since, both the IC 1848 clusters do not have early B-type stars reported in literature, we have considered only the strong feedback from the O-type stars here. If there are additional early B-type stars, identifying them would require spectroscopic observations, which is beyond the scope of this study.  We adopt the FUV luminosity (log (L$_\mathrm{FUV}$/L$_\odot$)) of each of the O-type stars (listed in section~\ref{sec:intro_w5clusters}) in both the clusters from \citet{guarcello2016}. Then the total FUV flux from the massive stars is calculated by incorporating the projected distance from each source to the massive stars. The fluxes are presented in terms of the Habing flux, G$_0$ (G$_0$=1.6 $\times$ 10$^{-3}$ ergs s$^{-1}$ cm$^{-2}$). In figure~\ref{fig:df_flux}, the dependence of disk fraction on the incident FUV flux is shown for sources above the mass completeness limit of 0.1 M$_\odot$. 

Here the bin size is fixed such that the number of diskless sources remains constant across all bins, so that the observed variation is a function of the number of disk sources. We do not find any significant variation in the disk population with respect to the incident radiation for both clusters. However, the overall fraction of the East cluster is marginally higher than the West cluster which can be attributed to the presence of a higher number of ionising sources in the latter.

%%%%%%%%%%%%%%%%%%%%%%%%%%%%%%%
%%%%%%%%%%%%%%%%%%%%%%%%%%%%%%%%

Recently, \citet{winter2022} reviewed the implications of external photoevaporation on the evolution of disks and consequentially on the theories of planet formation. They show that in massive star-forming regions with multiple O-type stars, where FUV flux is greater than 5000 G$_0$ (high UV environment), the disk rapidly erodes outside-in thereby interrupting the formation of giant planets in the outer disk. \citet{guarcello2007} investigated the effects of OB stars in NGC 6611 for a large area and they found that the UV radiation from these stars causes rapid photoevaporation of disks in their vicinity. This phenomenon of rapid disk destruction in such high radiation environments in the vicinity of OB stars has been reported in other massive regions as well such as NGC 2244, Arches, Cygnus OB2 (\citealt{balog2007,stolte2010,guarcello2016}). 

\begin{figure*}
    \centering
    \includegraphics[width=0.7\textwidth]{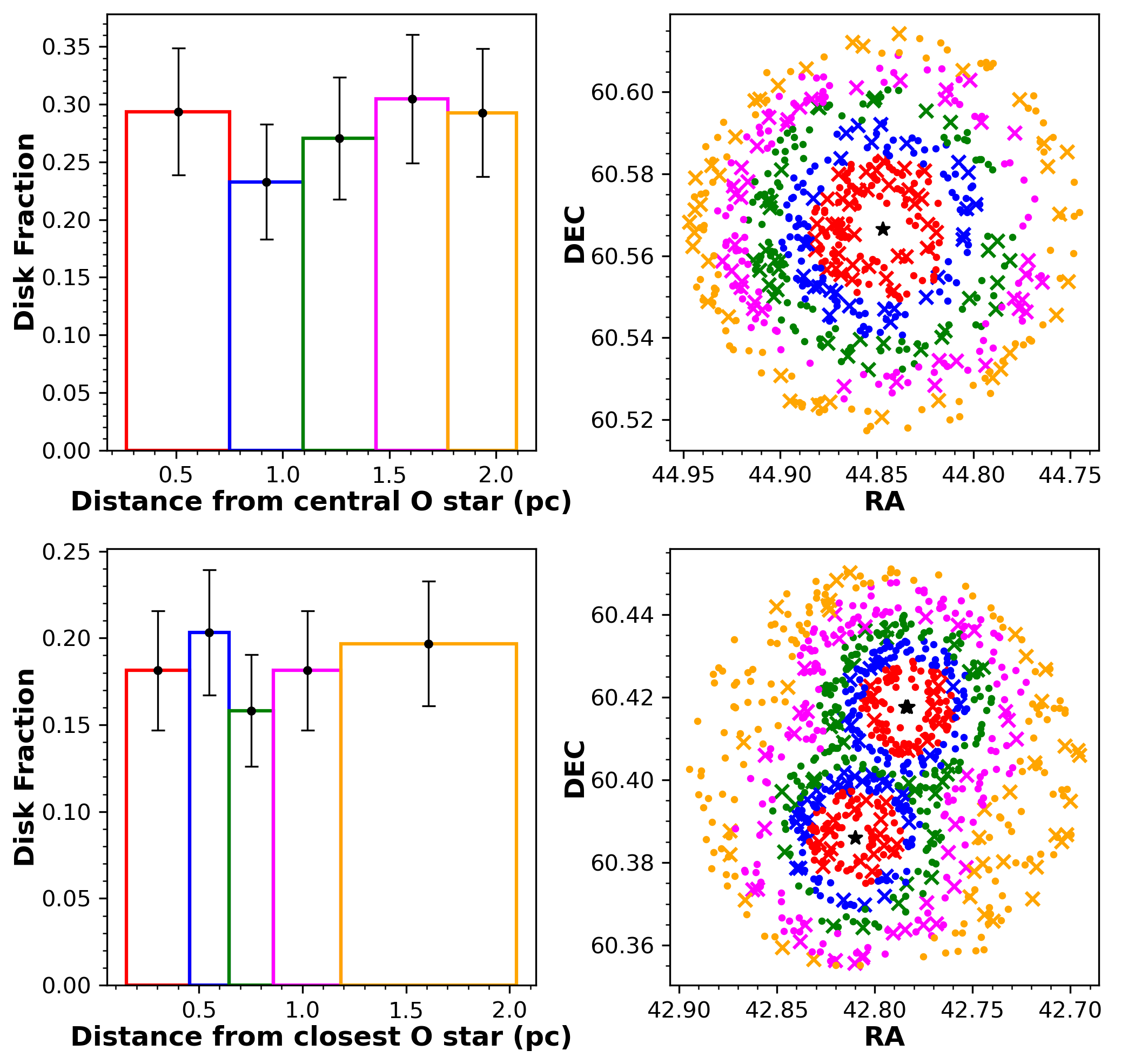}
    \caption{ (\textit{Top}) East cluster and (\textit{bottom}) West cluster. The disk fraction distribution as a function of the distance from the massive O-type star (left panel) and the spatial distribution of the disk (cross) and diskless (circle) sources within the 3$^\prime$ cluster area (right panel). The O-type stars in the region are marked with a black star. The binning is performed similarly as in figure~\ref{fig:df_flux} and the different colors in the right panel correspond to the sources in the respective color bins in the left panel. }
    \label{fig:df_spatial_dist}
\end{figure*}

We assessed the occurrence of disks for the completeness limited sample (above 0.1 M$_\odot$), radially spread across the cluster area from the central massive ionising sources. In the right panel of figure~\ref{fig:df_spatial_dist}, we show the spatial distribution of disk and diskless sources around the O-type stars as a function of their projected separation. The left panel shows a histogram representation of the same correlation as the right panel. We do not find any radial trend in the variation of disks present across the entire cluster area of $\sim$2 pc radius. As reported in \citet{allen2012}, the mixing of the stellar population due to the radial motion of the stars towards and away from the central ionising source could be a possible cause for the absence of significant spatial variation in the disk frequency. The decreased disk fraction in both clusters is plausibly due to the high ionising radiation experienced by the PMS sources, not just in their vicinity but also at the periphery of the cluster area under investigation. We discuss this further in section~\ref{sec:effects_radiation}.

%%%%%%%%%%%%%%%%%%%%%%%%%%%%%%%%%
%%%%%%%%%%%%%%%%%%%%%%%%%%%%%%%%%

We also analysed the disk color excess variation across the cluster area measured using the photometry from the NIR J band and the MIR IRAC [3.6], [4.5], [8.0] $\mu$m filters. We did not find any significant change in the disk excess away from the massive stars for the individual clusters as well as for the combined statistics of both clusters. Also, we do not observe any variation between the clusters. For the $\sigma$ Orionis cluster, \citet{mauco2023} using submm ALMA flux at 1.33 mm obtained the disk mass of sources in the region. They found that within 0.5 pc distance of the central massive star, $\sigma$ Ori, there is a lack of disks more massive than $\sim$3 M$_\oplus$. This shows that close to the ionising source, the disks have very low mass due to irradiation. However, while this study utilizing submm continuum flux probes the outer disk, our current investigation of the twin clusters explores the inner disk using NIR-MIR wavelengths. So we do not find any such drastic decrease in terms of the disk fraction and disk excess close to the cluster centers.

\subsection{Accretion properties}
\label{sec:accretion_properties}
Having determined the physical parameters of the PMS sources in the W5 clusters, we now constrain their accretion properties. The mass accretion rate parameter (M$_\mathrm{acc}$) quantifies the accretion of circumstellar gas onto the central protostar channeled by the magnetic field lines that act as accretion columns. The infalling material hits the photosphere at near free fall velocities producing accretion shocks which cause continuum excess emission predominantly seen in optical and UV wavelengths (\citealt{calvet1998,venuti2014}). A direct measure of the accretion rate is through measuring the UV excess or veiling of the photospheric lines. On the other hand, the accretion funnel flows produce characteristic emission lines, which can be used as tracers to measure the accretion rates (\citealt{white2003,kalari2021}). 

Here we follow the method proposed by \citet{demarchi2010} which uses broadband optical photometry along with the narrowband H$\alpha$ imaging to identify the accreting PMS sources. Then by using the empirical correlation between the accretion luminosity (L$_\mathrm{acc}$) and the H$\alpha$ line luminosity (L$_\mathrm{H\alpha}$), the L$_\mathrm{acc}$ parameter for the sources is determined from which the M$_\mathrm{acc}$ is derived. Photometric studies of accretion properties in the past have used the (r-H$\mathrm{\alpha}$) color as an indicator of the H$\alpha$ line strength relative to the r-band photospheric continuum (\citealt{barentsen2011,dutta2015,kalari2015,yadav2022}). The (r-H$\mathrm{\alpha}$) color excess is measured by comparing the observed (r-H$\mathrm{\alpha}$) color of the PMS source with the template photospheric (r-H$\mathrm{\alpha}$) color corresponding to the respective spectral type. Since main sequence stars do not have significant H$\mathrm{\alpha}$ in emission, we model their color as the photospheric template from \citet{drew2005}. Accreting stars with H$\mathrm{\alpha}$ in emission show higher (r-H$\mathrm{\alpha}$) color than other stars of the same spectral type. 

\begin{figure*}
    \centering
    \includegraphics[width=\columnwidth]{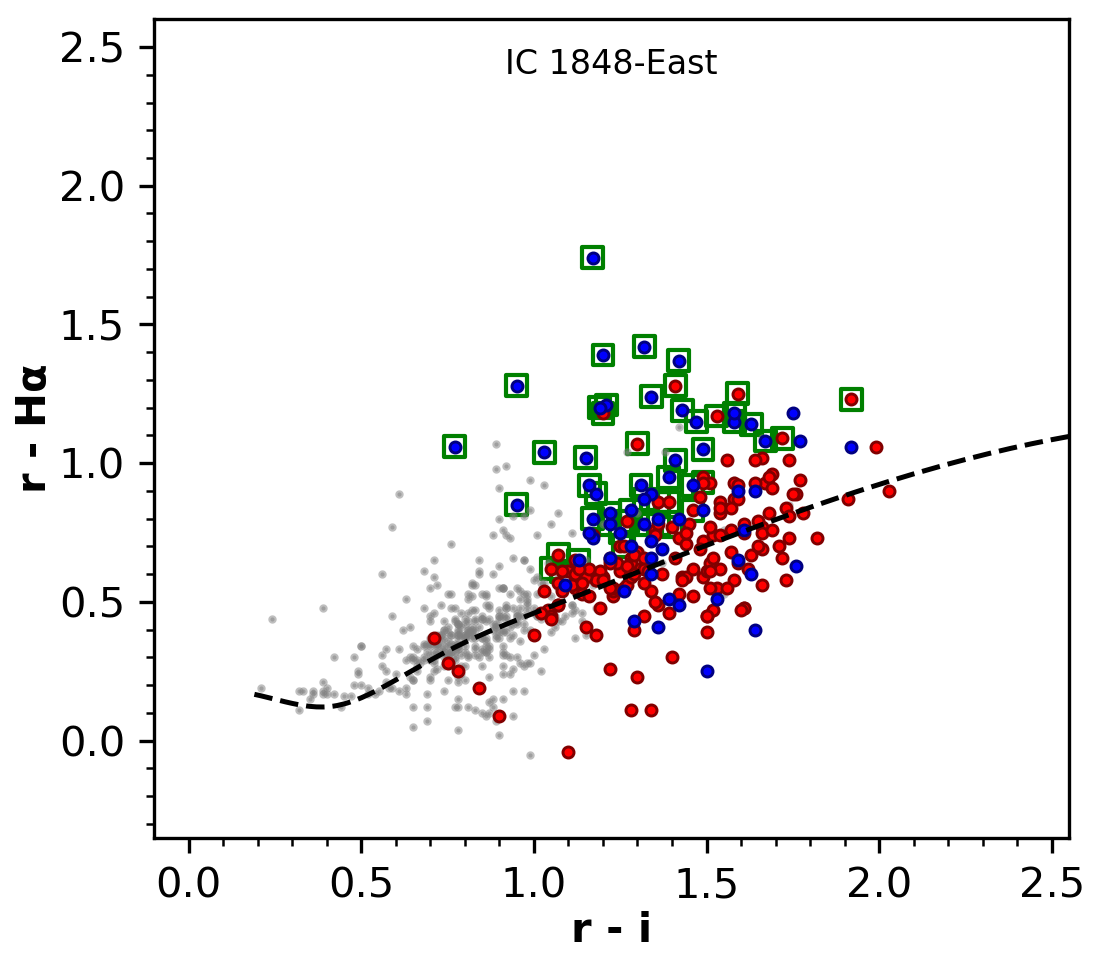}
    \includegraphics[width=\columnwidth]{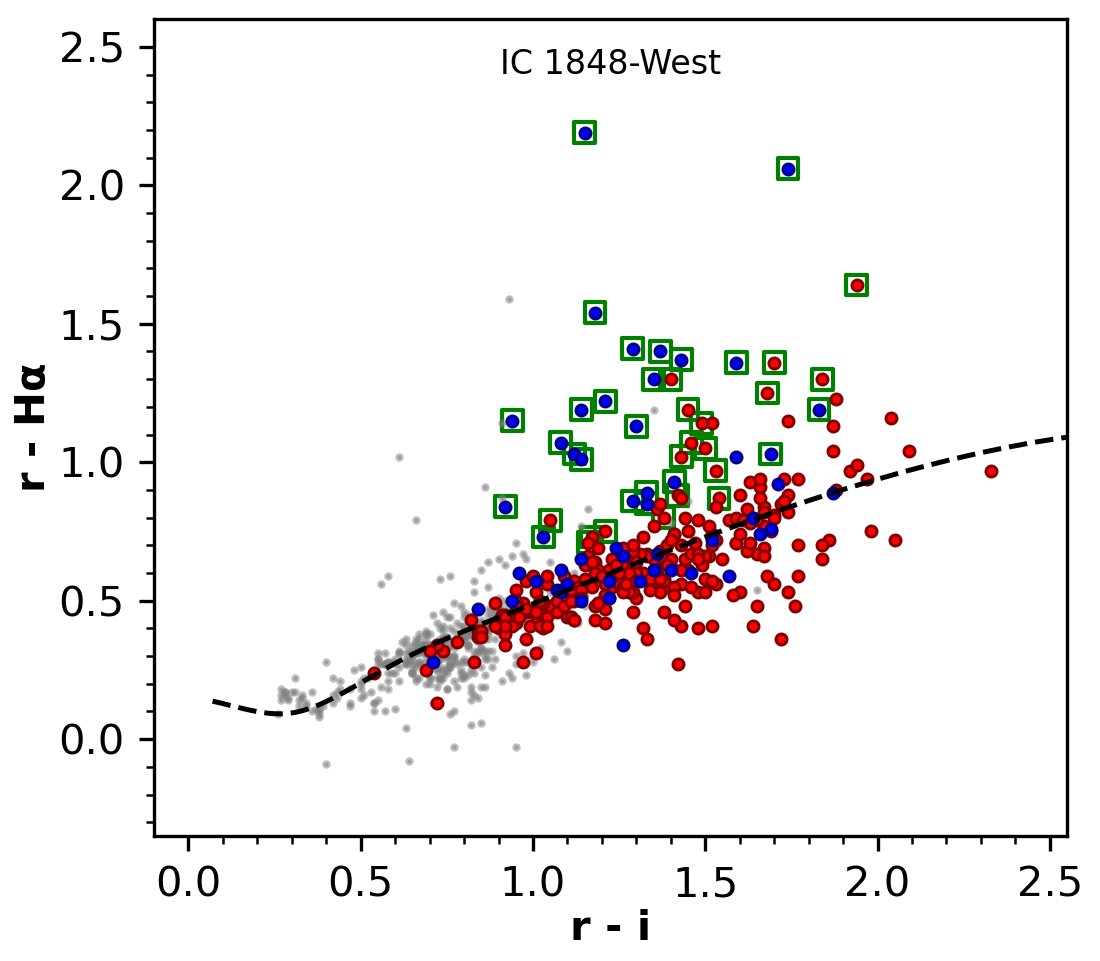}
    \caption{Color color diagram of IC 1848-East (left) and IC 1848-West (right) using IPHAS r, i, and H$\alpha$ bands. The grey dots are the sources within the cluster area with uncertainty $<$0.2 mag in all the three bands, the red and blue dots highlight the PMS non-excess and excess sources (age $<$ 10 Myr) and the green squares are the accreting sources identified in section~\ref{sec:Halpha_EW}. The black dashed curve is the reddened photospheric template from \citet{drew2005}.}
    \label{fig:ccd}
\end{figure*}

We make use of the IGAPS photometry in the r, i, and H$\alpha$ IPHAS filters to study the accretion properties. We select the PMS sources (age $<$10 Myr) with photometric uncertainty $<$0.2 mag in all three filters.  We note that the following analysis using H$\alpha$ photometry is limited by the sensitivity of the IPHAS survey. In figure~\ref{fig:ccd}, we show the (r-i) versus (r-H$\alpha$) color-color diagram for the East and West clusters. The (r-i) color is taken as the proxy for spectral type and the (r-H$\alpha$) color excess is given as,
\begin{equation}
    \mathrm{(r-H\alpha)_{excess}=(r-H\alpha)_{obs} - (r-H\alpha)_{model}}
\end{equation}

\noindent The reddened color of main sequence stars from \citet{drew2005} at a given (r-i) color is taken as the (r-H$\alpha$)$_\mathrm{model}$. The photospheric template is reddened by applying the mean extinction A$_\mathrm{K}$=0.21 and 0.14 mag for the East and West clusters respectively and using the extinction relations from \citet{cardelli1989} and \citet{barentsen2011}. We then considered the sources with (r-H$\alpha$)$_\mathrm{excess}$ exceeding at least three times the combined mean photometric uncertainties in the three bands \citep{carini2022} (i.e. $>$3$\delta$ where $\delta=\sqrt{\frac{\delta^{2}_{r}+\delta^{2}_{i}+\delta^{2}_{H\alpha}}{3}}$). This condition primarily selects only sources with reliable photometry in all the bands among which we further identify the accreting sources.

%%%%%%%%%%%%%%%%%%%%%%%%%%%%%%%
%%%%%%%%%%%%%%%%%%%%%%%%%%%%%%%%

\subsubsection{H$\alpha$ Equivalent Width (EW)}
\label{sec:Halpha_EW}
We measure the equivalent width (EW) of the H$\alpha$ line for sources with (r-H$\alpha$)$_\mathrm{excess}$ $>$ 3$\delta$ using the below equation,
\begin{equation}
    \label{eq:EW}
    \mathrm{EW = RECTW \times [1-10^{0.4\times(r-H\alpha)_{excess}}]}
\end{equation}

\noindent as shown by \citet{demarchi2010, kalari2015, tsilia2023, vlasblom2023}. RECTW is the rectangular bandwidth of the IPHAS H$\alpha$ filter (95 $\textup{\AA}$). In order to exclude older slow rotating stars and brown dwarfs with significant chromospheric activity, we select sources with EW $<$-10 $\textup{\AA}$ \citep{white2003} as candidate H$\alpha$ emitting accreting sources. By applying both these conditions, we identify 51 and 41 accreting sources in the East and West clusters, respectively which are shown in figure~\ref{fig:ccd}. In both clusters, some of the previously identified non-excess sources exhibit H$\alpha$ excess and have been identified as accreting sources. A similar phenomenon has been recently reported in the Lagoon nebula by \citet{venuti2024}.  For these 92 sources (among which 56 are excess sources), we further derive the accretion luminosity and mass accretion rates as detailed in the following sections. 

\subsubsection{H$\alpha$ Line Luminosity (L$_{H\alpha}$)}
The total flux in the H$\alpha$ band (F$_\mathrm{total}$) is the sum of the H$\alpha$ line flux (F$_\mathrm{H\alpha}$) and the stellar continuum flux (F$_\mathrm{continuum}$). The total H$\alpha$ flux can be measured from the dereddened H$\alpha$ magnitude using the relation,
\begin{equation}
    F_{total} = F_{0} \times [10^{-0.4(m_{H\alpha}+0.03)}]
\end{equation}

\noindent where m$_\mathrm{H\alpha}$ is the dereddened H$\alpha$ magnitude, F$_{0}$ is the H$\alpha$ band-integrated reference flux = 1.72$\times$10$^{-7}$ ergs s$^{-1}$ cm$^{-2}$ that corresponds to the magnitude of Vega, for which the m$_\mathrm{H\alpha}$ is taken as 0.03 mag (refer~\citealt{barentsen2011}). The line flux (F$_\mathrm{H\alpha}$) can be obtained from the total flux using the EW calculated in \eqref{eq:EW},
\begin{flalign}
    F_{H\alpha} & =  F_{total} - F_{continuum}\\
    & =  F_{total} \times \frac{-EW/RECTW}{1-EW/RECTW}
\end{flalign}

\noindent We then derive the H$\alpha$ line luminosity (L$_\mathrm{H\alpha}$) as follows,
\begin{equation}
    L_{H\alpha} = F_{H\alpha} \times 4\pi d^{2}
\end{equation}

\noindent where d is the mean distance of the East (2380$\pm$510 pc) and West (2220$\pm$420 pc) clusters. With the estimated H$\alpha$ line luminosity, we can derive the accretion luminosity of the sources as discussed below.

\subsubsection{Accretion Luminosity (L$_{acc}$)}
To obtain the accretion luminosity (L$_\mathrm{acc}$), we use the empirical relation between L$_\mathrm{H\alpha}$ and L$_\mathrm{acc}$ given by \citet{barentsen2011} as,
\begin{equation}
    log L_{acc} = (1.13\pm0.07)log L_{H\alpha} + (1.93\pm0.23)
\end{equation}

\begin{figure}
    \centering
    \includegraphics[width=\columnwidth]{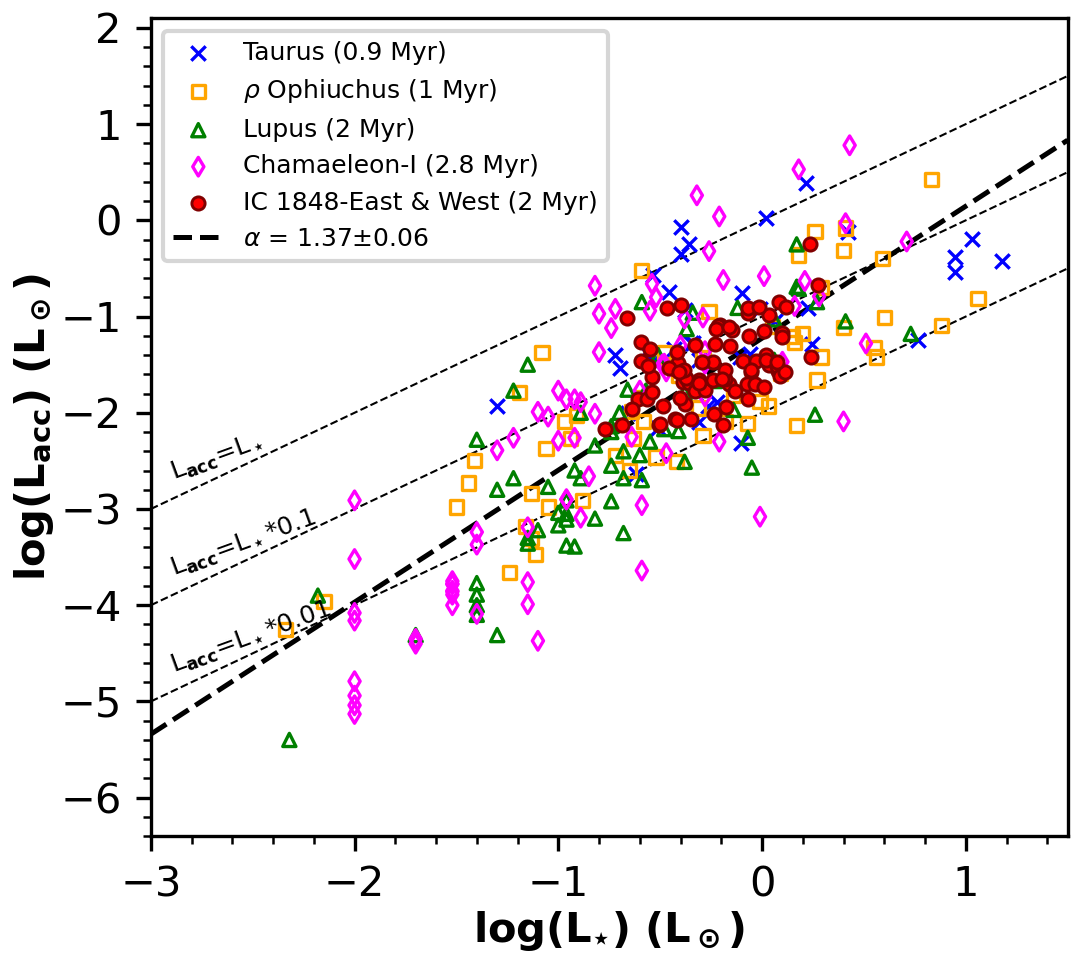}
    \caption{Accretion luminosity as a function of stellar luminosity. The colored markers denote the population of accretors from different regions as indicated in the legend. The dashed black line shows the linear best fit of the combined distribution of the sources from all the regions presented here with a slope of 1.37$\pm$0.06. The dotted lines represent the linear relationship between L$_\mathrm{acc}$ and L$_\star$ for varying proportionality constants as indicated. }
    \label{fig:lum_lacc}
\end{figure}

Figure~\ref{fig:lum_lacc} shows the distribution of the sources in the L$_\mathrm{acc}$- L$_{\star}$ plane. Due to the low statistics of accreting sources in the individual clusters, in the following analysis of the accretion rates we have combined the population in both the clusters. For the twin clusters, given the limited mass range between $\sim$0.2-1.3 M$_\odot$, the accretion luminosity (log L$_\mathrm{acc}$) varies between -0.2 to -2.2 L$_{\odot}$. As already reported in other star-forming regions, L$_\mathrm{acc}$ is observed to increase with L$_{\star}$.  With the limited luminosity range of the sources in the twin clusters, it is difficult to trace its correlation against the accretion luminosity. So we compare our results with the distribution of the similar age star-forming regions like Taurus ($\sim$ 0.9 Myr), $\rho$ Ophiuchus ($\sim$ 1 Myr), Lupus ($\sim$ 2 Myr), and Chamaeleon-I ($\sim$ 2.8 Myr) taken from previous surveys by \citet{gangi2022} and \citet{testi2022}. In these studies, the accretion luminosity is derived by estimating the line luminosity of different spectral emission lines. The combined linear fit to the distribution of sources taken from literature along with the sources in IC 1848-East and West gives a slope of 1.37$\pm$0.06 which is consistent with the slope estimated for the various star-forming regions (\citealt{alcala2014,manara2017,manara2021,gangi2022,abad2024}). \citet{wichittanakom2020} compiled the best fit for a group of classical T Tauri stars and reported the correlation as L$_\mathrm{acc}$ $\propto$ L$_\star^{1.17\pm0.09}$ which is comparable to our results for the similar luminosity range. Additionally, we found that deriving the correlation between L$_\mathrm{acc}$ and L$_\star$ excluding the sources in IC 1848-East and West produced a slope of 1.39$\pm$0.07. This indicates that the correlation is primarily driven by the members of other star forming regions and the accreting sources in the twin clusters follow a similar distribution.

The lines of constant L$_\mathrm{acc}$/L$_{\star}$ ratio are overplotted in Figure~\ref{fig:lum_lacc} for increasing coefficient values from 0.01 to 1. We observe that the majority of the objects have L$_\mathrm{acc}$ $<$ 0.1 $\times$ L$_\star$. This distribution has been previously reported in similar studies by \citet{natta2006}, \citet{rigliaco2011}, and \citet{biazzo2014}, although there is a significant fraction of sources between L$_\mathrm{acc}$ $=$ L$_\star$ and L$_\mathrm{acc}$ $>$ 0.1 $\times$ L$_\star$. Additionally, we observe that among the sources belonging to the W5 clusters, the spread in L$_\mathrm{acc}$ is narrow compared to the scatter in the distribution seen for other regions. The possible cause for this relatively narrow spread could be the confined age of the cluster members and very low extinction in the region. However, we also note that this may be an effect of the completeness of the IPHAS data used for the study.

%%%%%%%%%%%%%%%%%%%%%%%%%%%%%%%
%%%%%%%%%%%%%%%%%%%%%%%%%%%%%%%%

For sources with low accretion rates, the contribution due to the chromospheric emission may impact the estimates of L$_\mathrm{acc}$. This contribution was derived as the noise in the accretion luminosity (L$_\mathrm{acc,noise}$) by \citet{manara2013} and \citet{manara2017chrom} as log (L$_\mathrm{acc,noise}$/L$_\star$) $=$ (6.2 $\pm$ 0.5) log T$_\mathrm{eff}$ - (24.5 $\pm$ 1.9) for T$_\mathrm{eff}$ $\leq$ 4000 K and log (L$_\mathrm{acc,noise}$/L$_\star$) $=$ -2.3 $\pm$ 0.1 for 4000 K $<$ T$_\mathrm{eff}$ $<$ 5800 K, respectively. This relation is taken as the locus below which the chromospheric contribution is significant. We find that all our accreting sources in both the clusters fall well above this limit so we neglect the contribution from the chromosphere for our sources.

\subsubsection{Mass accretion rate}
Having estimated the accretion luminosity, we now derive the mass accretion rate (M$_\mathrm{acc}$) using the same relation from \citet{hartmann1998} adopted by previous studies in the literature. 
\begin{equation}
    L_{acc} = \frac{GM_{\star}M_{acc}}{R_{\star}}(1-\frac{R_\star}{R_{in}})
\end{equation}

\noindent where, stellar radius (R$_\star$) and stellar mass (M$_\star$) were estimated in section~\ref{sec:sed_vosa} based on the SED analysis. G is the gravitational constant and we adopt the standard value for the inner radius R$_\mathrm{in}$ $=$ 5R$_\star$ (\citealt{gullbring1998}). Therefore,
\begin{equation}
    M_{acc} = 1.25 \frac{L_{acc}R_\star}{GM_\star}
\end{equation}

The collective accretion rate of both the East and West clusters ranges between  5.8$\times$10$^{-10}$ to 9.3$\times$10$^{-8}$ M$_\odot$/yr with a mean mass accretion rate 6.4 $\times$ 10$^{-9}$ M$_\odot$/yr for a mass range of 0.2 to 1.3 M$_\odot$. Considering our criterion of EW $<$ -10 $\textup{\AA}$ to select accreting sources, for a source of H$\alpha$ = 18 mag, this corresponds to an accretion luminosity (log L$\mathrm{_{acc}}$) of -1.7 and mass accretion rate (M$\mathrm{_{acc}}$) of 1.8$\times$10$^{-9}$ M$_\odot$/yr which would indicate an approximate upper limit on the accretion rates. 

The contribution of error in the derived parameters can be attributed to, (i) the uncertainty in the photometry of IGAPS filters, (ii) error in the estimates of stellar radius and mass, (iii) scatter in the distances of individual sources compared to the adopted uniform value, (iv) scatter in the adopted conversion relation of L$_\mathrm{H\alpha}$ to L$_\mathrm{acc}$. All these errors are propagated through the estimation of each parameter. 

Previously, \citet{lim2014} obtained the mass accretion rate for a population of PMS stars in the IC 1848-West cluster using UV-excess and found the mean to be 1.4 $\times$10$^{-8}$ M$_\odot$/yr in the mass range 0.5 to 2 M$_\odot$.

\subsubsection{Accretion rates as a function of mass and age}
The relation between the mass accretion rates and the stellar mass for various star-forming regions in the Milky Way as well as in the LMC and SMC have shown to be steeper than linear. In recent years with more and more spectroscopic surveys, the slope of the relation has been reported to be $\sim$1.6-2.3 with a large spread in M$_\mathrm{acc}$ at any given M$_\star$ of about $\sim$1-2 dex (\citealt{alcala2014,venuti2014,hartmann2016,manara2016,alcala2017,manara2017,venuti2019}).  

\begin{figure}
    \centering
    \includegraphics[width=\columnwidth]{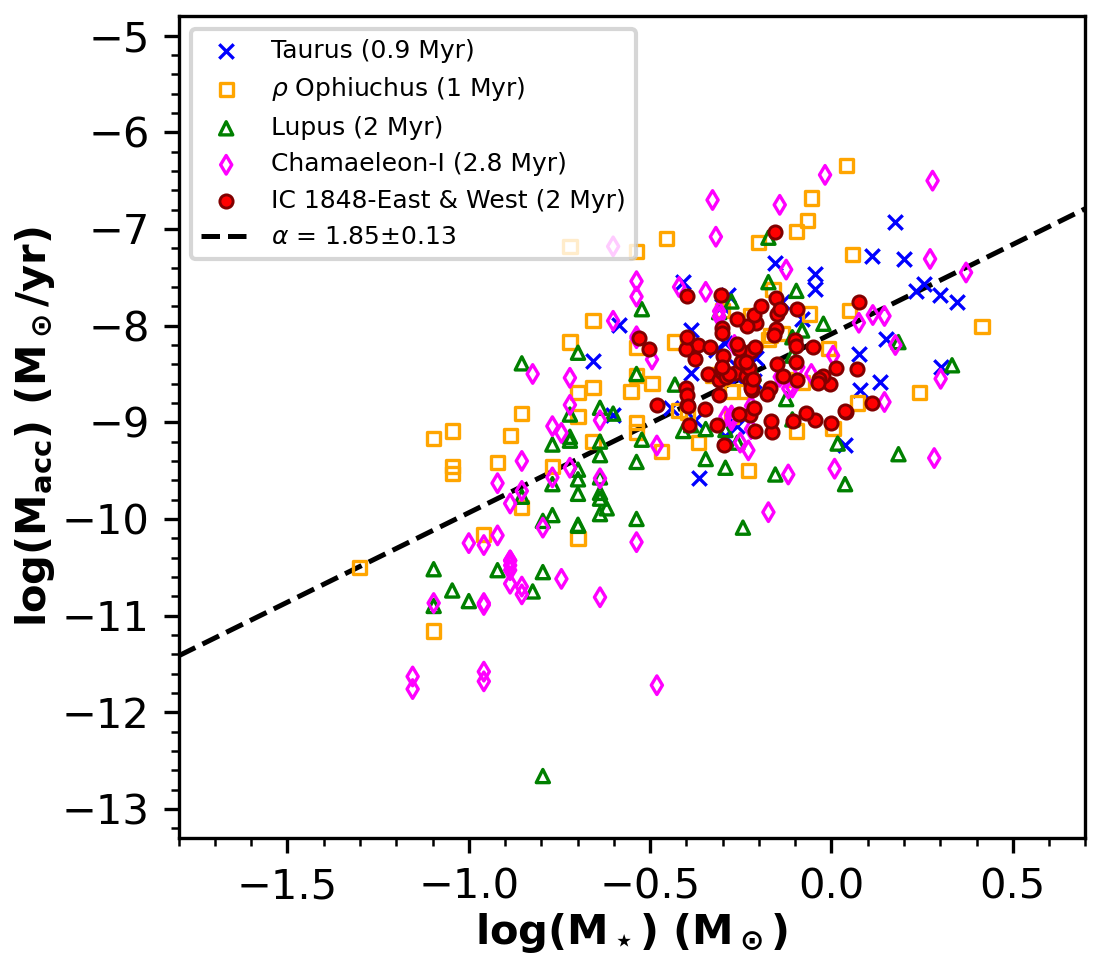}
    \caption{Mass accretion rate as a function of stellar mass for targets in various star-forming regions including IC 1848-East and West clusters. The color scheme is the same as figure~\ref{fig:lum_lacc} and shown in legend. }
    \label{fig:mass_macc}
\end{figure}

Figure~\ref{fig:mass_macc} shows the distribution of M$_\star$ vs M$_\mathrm{acc}$ for the W5 twin clusters along with some of the similar age star-forming regions in the solar neighbourhood. The best fit to the combined distribution of sources of all the regions presented in the plot has a slope $\alpha$ $=$ 1.85$\pm$0.13. At any given M$_\star$, there is a large dispersion in the M$_\mathrm{acc}$ values, of about one to two orders of magnitude, which has been observed in most such studies suggesting that the spread is physical and not due to observational biases (\citealt{betti2023,manara2023}).

%%%%%%%%%%%%%%%%%%%%%%%%%%%%%%%
%%%%%%%%%%%%%%%%%%%%%%%%%%%%%%%%

\begin{figure}
    \centering
    \includegraphics[width=\columnwidth]{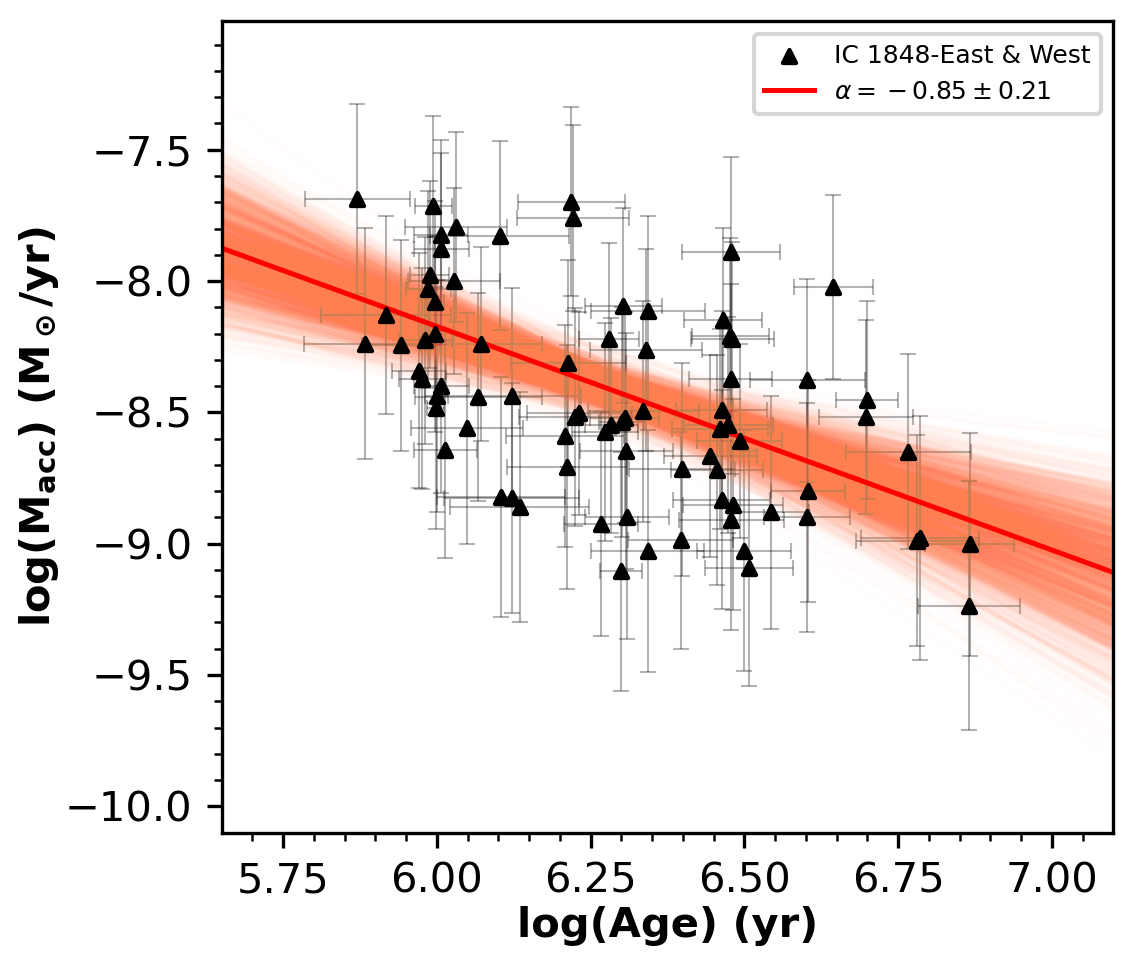}
    \caption{Mass accretion rate as a function of stellar age for the sources in the twin clusters. The error bars mark their corresponding uncertainty in both axes. The linear fit from a few of the multiple regressions is shown here in orange and the best fit distribution using the average of the parameters of all the iterations is marked in red.} 
    \label{fig:age_macc}
\end{figure}

Figure~\ref{fig:age_macc} shows the distribution of the mass accretion rate of the W5 clusters as a function of the age of the sources estimated in section~\ref{sec:sed_vosa}. To estimate the linear correlation, we use the method from \citet{kelly2007} incorporating the uncertainties in both the M$_\mathrm{acc}$ and the age of the sources. We perform around 5000 iterations and obtain the average slope and standard deviation as -0.85$\pm$0.21 which is consistent with the slope reported in literature  (\citet{hartmann2016} and references therein). However, due to the spread in the estimated age and accretion rates, this relation between the two parameters must be treated with caution. Large samples of accretion rates across regions, spanning a large range of truly different ages, are essential before conclusions may be drawn on the relationship.

%%%%%%%%%%%%%%%%%%%%%%%%%%%%%%%
%%%%%%%%%%%%%%%%%%%%%%%%%%%%%%%%

\section{Discussion}
\label{sec:discussion}
\subsection{Effect of external photoevaporation in the twin clusters}
\label{sec:effects_radiation}
The influence of environment on disk properties such as truncation in disk radii, lower disk mass, and shortened disk lifetime have been previously reported in star-forming regions of high UV radiation environment 
(\citealt{ansdell2017,eisner2018,terwisga2020}). Recently, \citet{terwisga2023} studied the disks around Class II YSOs in the ONC and showed that even in intermediate FUV radiation fields (10-1000 G$_0$) the effect of environment is significant. For their sample of YSOs they found the median disk dust mass reduces by half for two orders of magnitude increase in the FUV field strength over the lifetime of the region. 

\citet{haworth2018,haworth2023} modeled the disk mass loss rate for externally irradiated disks at various star formation environments. Based on their models, UV field strengths of $>$10$^3$ G$_0$ can drive strong winds inducing a mass loss rate of above 10$^{-6}$ M$_{\odot}$/yr. As can be seen from figure~\ref{fig:df_flux}, the YSOs in the twin clusters are exposed to a FUV radiation field higher than $\sim$10$^3$ G$_0$. This shows that the entire core of the IC 1848 clusters is under the influence of high radiation feedback from the environment. Additionally, we note that the clusters evolve under the expulsion of gas due to the radiation and stellar winds from the central massive stars, which makes it difficult to estimate the temporally integrated exposure of the disks to UV radiation (\citealt{karnath2019,megeath2022}). In a recent study by \citet{coleman2022}, they traced the evolution and dispersal time scales of disks under both internal and external photoevaporation and in different environments. They show that the disk fraction is affected by the strong stellar winds in the vicinity of massive stars and systems experiencing mass loss rates around 10$^{-6}$ M$_{\odot}$/yr have a low disk fraction ($\sim$20\% at 2 Myr). This is consistent with our findings for the twin clusters which accentuates the understanding that the environment plays a significant role in faster disk dispersal and in turn, affects the planet formation.

\subsection{Importance and implication of the twin clusters}
As outlined in section~\ref{sec:intro_w5clusters}, the environmental effects on low-mass star formation have not been extensively analysed in distant massive regions limited by the high extinction and sensitivity to detect these objects in crowded regions. However, the IC 1848-East and West clusters within the W5 complex make exclusive test beds to carry out such studies. They are considered to be comparable clusters owing to their similar distance ($\sim$2 kpc), age ($\sim$2 Myr), low extinction (A$_\mathrm{V}$ $\sim$1.5 mag) and peak mass ($\sim$ 0.2-0.3 M$_\odot$) but differ in stellar density (by a factor of two) and FUV field strength (by a factor of four). Our deep broad spectrum photometry in optical, NIR, and MIR wavelengths, enables the detection of low-mass YSOs down to the substellar regime ($\sim$0.01 M$_\odot$) in both clusters. Our data is $>$80\% complete in a minimum of five IR bands down to $\sim$0.07 M$_\odot$. As can be seen from the age-mass distribution (refer figure~\ref{fig:hrd}) and various CMDs (refer figure~\ref{fig:cmds}), both the clusters exhibit akin characteristics.

We studied the role of external radiation as well as the influence of stellar mass on the presence of protoplanetary disks in both clusters. Considering the young age of the clusters, we find the disk frequency to be lower ($\sim$20\%) than that observed in nearby regions of similar age. This decrease is attributed to the strong UV radiation field throughout the area within $\sim$2 pc radius, caused by the massive O-type stars present at the center of both clusters. Although there is a moderate decline in the overall fraction of disk sources associated with the massive star population in both clusters, this decrease is more pronounced under the higher FUV field. We note that this association of decreased disk population with the FUV radiation is considered relying on the estimated average age of the clusters, which is dependent on the evolutionary models used.

This is a one-of-a-kind study carried out on two distant regions using similar deep multi-band data complete down to the stellar-substellar boundary owing to the uniform minimal extinction in the region. In both the clusters the YSOs are distributed within a compact area ($\sim$2 pc radius) and exhibit a well defined distribution along the location of the PMS branch in CMDs and hence our catalog is affected by limited contamination from background sources. Located beyond the solar neighbourhood, these clusters can be used as reference templates for future studies of outer galaxy regions. Additionally in the stellar density spectrum, these clusters lie in between low density regions like Taurus, 25 Ori (\citealt{esplin2019,suarez2019}) and high density supermassive clusters like Arches, RCW 38, NGC 3603 (\citealt{beccari2010,stolte2010,muzic2017}). Furthermore, these twin clusters can be used to validate and refine theoretical models and simulations to test environmental effects on disk evolution, ensuring that the models accurately represent the complexities of real-world star-forming environments. Likewise, investigating feedback mechanisms, such as stellar winds and radiation pressure, within each cluster can provide insights into how these mechanisms vary based on environmental factors and influence subsequent star and planetary formation.

\section{Conclusion}
\label{sec:conclusion}
 In this study, we use deep multiband optical, NIR and MIR photometry to explore the influence of massive stars on low-mass star formation in two distant clusters (IC 1848-East and West) in the W5 complex. These clusters are unique test beds having similar (in terms of distance, age, and extinction) yet distinct (in terms of density and UV radiation fields) characteristics. We studied the core of both clusters (within 3$^\prime$ radius) to understand the role of environment in disk evolution. The key results of the study are briefed below,

\begin{enumerate}
    \item We obtain the physical parameters of the sources by building their SEDs and the PMS sources are identified based on their location in the HR diagram. 
    \item The frequency of disk bearing sources in the East and West clusters was determined to be $\sim$27$\pm$2\% and  $\sim$17$\pm$1\%, respectively. We observe the disk fraction in both clusters to be lower than other nearby young star forming regions, with an overall decrease in the disk frequency among the members of the West cluster than the East cluster across all masses. 
    \item The effect of FUV radiation from the central massive stars was quantified and we found no significant variation in the disk population as a function of the incident radiation for both clusters. However, the overall decrease in the disk fraction given the young age of the clusters can be attributed to the entire region being under the influence of high ionising UV radiation.
    \item We also assessed the occurrence of disks radially spread across the cluster area and their distribution with respect to the distance from the ionising stars. There is no radial trend in the disk population within the cluster area of $\sim$2 pc. 
    \item Using  H$\alpha$ photometry the mass accretion rates were derived to range between 5.8$\times$10$^{-10}$ to 9.3$\times$10$^{-8}$ M$_\odot$/yr with a mean mass accretion rate 6.4 $\times$ 10$^{-9}$ M$_\odot$/yr for a mass range of 0.2 to 1.3 M$_\odot$ for the combined statistics of the twin clusters. The accretion rates correlate with the stellar mass and age by a slope of 1.85$\pm$0.13 and -0.85$\pm$0.21, respectively and both the slope values are consistent within the range reported in the literature for other young regions.
\end{enumerate}

Comparing our observational results to theoretical models in literature we see that the entire core of the cluster under investigation experiences high UV radiation above 10$^3$ G$_0$ from the environment. This high radiation drives strong disk winds causing high disk mass loss rates consistent with the reduced disk fraction observed in the young twin clusters. Our results show that the environment plays a significant role in the evolution and dissipation of disks, understanding which is important in planet formation studies. We highlight that these twin clusters, owing to the properties discussed above can be utilised as critical references to advance our understanding of star formation, stellar evolution, and the role of environmental factors in shaping these processes.

\section*{Acknowledgements}
This publication made use of the data products from the Two Micron All Sky Survey (a joint
project of the University of Massachusetts and the Infrared Processing and Analysis Center/California Institute of Technology, funded by NASA and NSF), archival data obtained with the Spitzer Space Telescope (operated by the Jet Propulsion Laboratory, California Institute of Technology, under a contract with NASA), and the NOAO Science archive, which is operated by the Association of Universities for Research in Astronomy (AURA), Inc., under a cooperative agreement with the National Science Foundation. Based on observations obtained at the 3.6m Devasthal Optical Telescope (DOT), which is a National Facility run and managed by Aryabhatta Research Institute of Observational Sciences (ARIES), an autonomous Institute under Department of Science and Technology, Government of India and observations obtained with MegaPrime/MegaCam, a joint project of CFHT and CEA/DAPNIA, at the Canada-France-Hawaii Telescope (CFHT) which is operated by the National Research Council (NRC) of Canada, the Institut National des Science de l'Univers of the Centre National de la Recherche Scientifique (CNRS) of France, and the University of Hawaii. The observations at the Canada-France-Hawaii Telescope were performed with care and respect from the summit of Maunakea which is a significant cultural and historic site. This publication makes use of VOSA, developed under the Spanish Virtual Observatory project supported by the Spanish MINECO through grant AyA2017-84089. VOSA has been partially updated by using funding from the European Union's Horizon 2020 Research and Innovation Programme, under Grant Agreement n$^\circ$ 776403 (EXOPLANETS-A). BD is thankful to the Center for Research, CHRIST (Deemed to be University), Bangalore, India. JJ acknowledges the financial support received through the DST-SERB grant SPG/2021/003850. SRD acknowledges support from the Fondecyt Postdoctoral fellowship (project code 3220162) and ANID BASAL project FB210003. DKO acknowledges the support of the Department of Atomic Energy, Government of India, under Project Identification No. RTI 4002. The authors also thank and acknowledge Tom Megeath for his insightful feedback on this paper.

%%%%%%%%%%%%%%%%%%%%%%%%%%%%%%%%%%%%%%%%%%%%%%%%%%
\section*{Data Availability}
The data underlying this article will be shared on reasonable request to the corresponding author.



\bibliographystyle{mnras}
\bibliography{main} % if your bibtex file is called example.bib

%\begin{thebibliography}{99}
%\end{thebibliography}

%%%%%%%%%%%%%%%%%%%%%%%%%%%%%%%%%%%%%%%%%%%%%%%%%%

%%%%%%%%%%%%%%%%% APPENDICES %%%%%%%%%%%%%%%%%%%%%

\appendix
\label{sec:appendix}

\section{Data completeness}
\begin{figure*}
    \centering
    \includegraphics[width=\columnwidth]{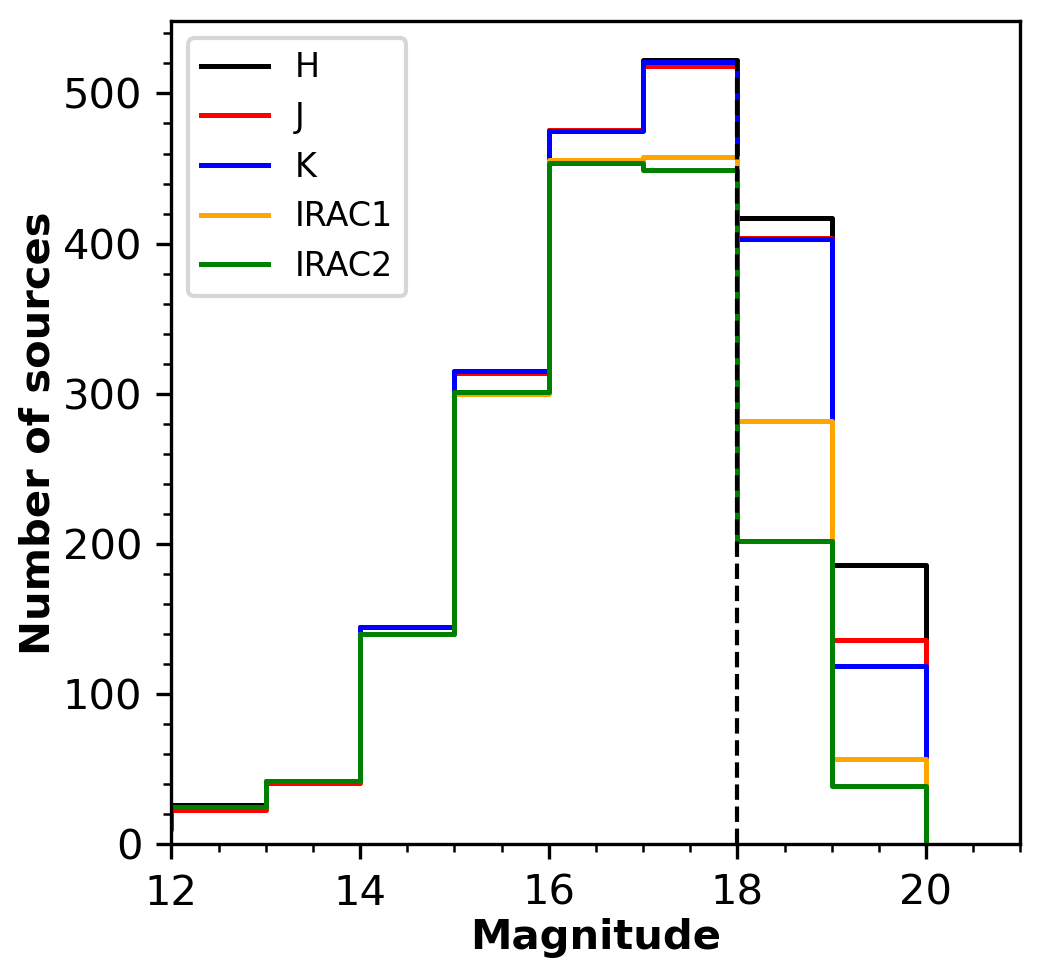}
    \includegraphics[width=\columnwidth]{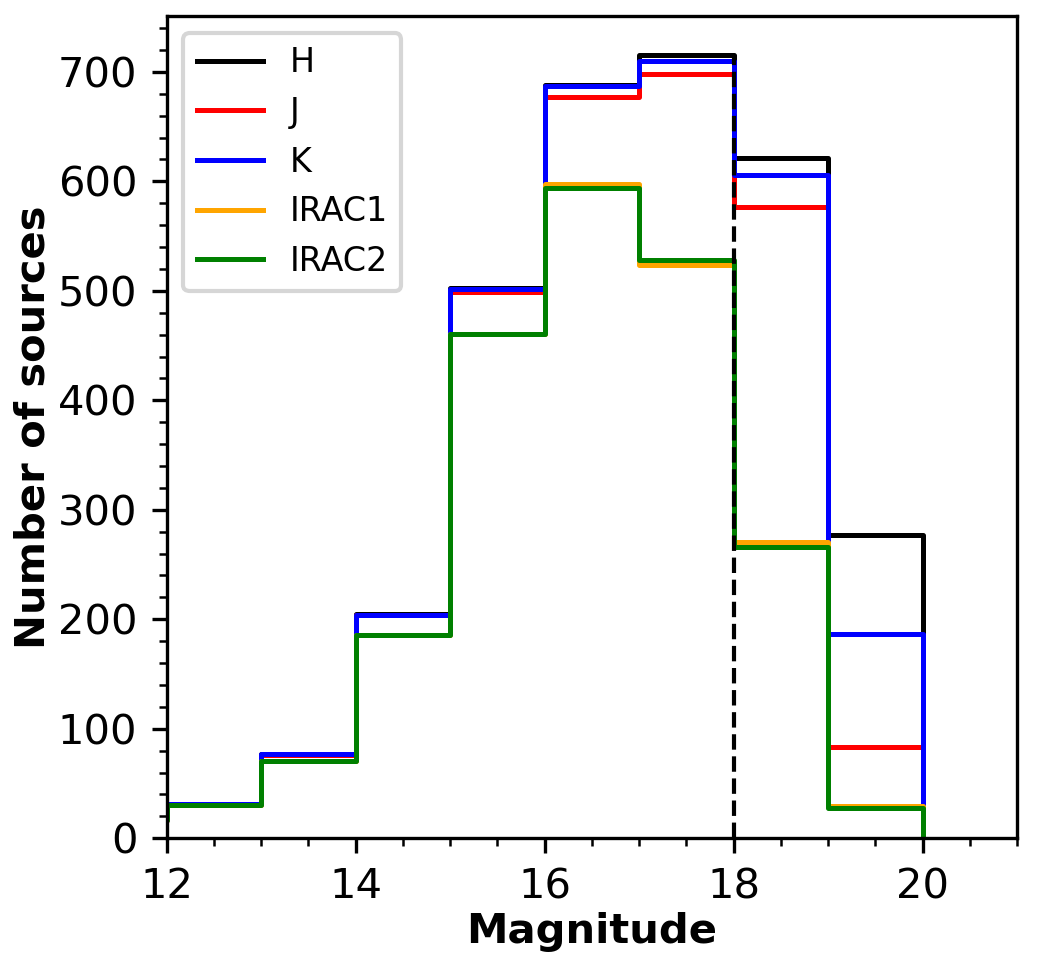}
    \caption{Histograms showing the completeness limits of the data in different filters as indicated by the legend for the East (\textit{left}) and West (\textit{right}) clusters with respect to the H band. The turnover point in the distribution serves as a proxy for the 90\% completeness limit of the H band and is marked by the vertical dashed line.}
    \label{fig:completeness}
\end{figure*}

In Figure~\ref{fig:completeness} the completeness of the data in the NEWFIRM J, K, Spitzer IRAC1 and IRAC2 bands relative to the  H band is shown as histograms. We consider the turnover point in the H band distribution at 18 mag as proxy for the 90\% completeness of the data.

%%%%%%%%%%%%%%%%%%%%%%%%%%%%%%%%%%%%%%%%%%%%%%%%%%

% Don't change these lines
\bsp	% typesetting comment
\label{lastpage}
\end{document}